\documentclass[useAMS,usenatbib]{mnras}

\usepackage{graphicx}
\usepackage{rotating}
\usepackage{longtable}
\usepackage{subfig}
\usepackage{float}

\newif\ifAMStwofonts                        
\newcommand{\lsimeq}{{_<\atop^{\sim}}}

\title[Inner morphology of core regions of NGC5135]{Unveiling the inner morphology and gas kinematics of NGC~5135 with ALMA}
\author[G. Sabatini et al.]
{\parbox{\textwidth}{\raggedright G. Sabatini$^{1, 2}$\thanks{E-mail: sabatini@ira.inaf.it \textit{or} giovanni.sabatini4@unibo.it},
C. Gruppioni$^{3}$,
M. Massardi$^{2}$,
A. Giannetti$^{2}$,
S. Burkutean$^{2}$,
A. Cimatti$^{1}$,
F. Pozzi$^{1}$,
M. Talia$^{1}$
 }
\vspace{0.4cm}\\
$^{1}$Dipartimento di Fisica e Astronomia, Universit\'a  degli Studi di Bologna, via Gobetti 93/2, I-40129 Bologna, Italy\\
$^{2}$INAF, Istituto di Radioastronomia - Italian ARC, via Gobetti 101, I-40129 Bologna, Italy\\
$^{3}$INAF, Osservatorio Astronomico di Bologna, via Gobetti 93/3, I-40129 Bologna, Italy\\
}
\begin{document}

\date{Accepted ???. Received 2017 ???; in original form 2017???}

\pagerange{\pageref{firstpage}--\pageref{lastpage}} \pubyear{2017}

\maketitle

\label{firstpage}

\begin{abstract}
The local Seyfert 2 galaxy NGC5135, thanks to its almost face-on appearance, a bulge overdensity of stars, the presence of a large-scale bar, an AGN and a Supernova Remnant, is an excellent target to investigate the dynamics of inflows, outflows, star formation and AGN feedback. Here we present a reconstruction of the gas morphology and kinematics in the inner regions of this galaxy, based on the analysis of Atacama Large Millimeter Array (ALMA) archival data. To our purpose, we combine the available $\sim$100 pc resolution ALMA 1.3 and 0.45 mm observations of dust continuum emission, the spectroscopic maps of two transitions of the CO molecule (tracer of molecular mass in star forming and nuclear regions), and of the CS molecule (tracer of the dense star forming regions) with the outcome of the SED decomposition. By applying the {\tt$^{\rm 3D}$BAROLO} software (3D-Based Analysis of Rotating Object via Line Observations), we have been able to fit the galaxy rotation curves reconstructing a 3D tilted-ring model of the disk. Most of the observed emitting features are described by our kinematic model. We also attempt 
an interpretation for the emission in few regions that the axisymmetric model fails to reproduce. The most relevant of these is a region at the northern edge of the inner bar, where multiple velocity components overlap, as a possible consequence of the expansion of a super-bubble.  
\end{abstract}

\begin{keywords}
galaxies: active -- galaxies: ISM -- galaxies: kinematics and dynamics -- galaxies: nuclei -- galaxies: Seyfert -- submillimetre: galaxies.
\end{keywords}

\section{Introduction}
\label{sec_intro}
Active galactic nuclei (AGN) are thought to play a major role in the formation and evolution of galaxies, providing mechanisms for feedback from the supermassive black hole (SMBH) to its host galaxy and the intergalactic medium (see, e.g., \citealt{fabian2012,somerville2015} and references therein). 
In co-evolutionary scenarios (e.g., \citealt{hopkins2007, lapi2014}) star formation (SF) activity and SMBH properties are closely connected, both in high redshift quasars and in local Seyfert nuclei, 
which are fuelled by accretion of material onto the SMBH. In the absence of major merging events and companions, as it is the case for the vast majority of local galaxies, the mechanisms that link SF and accretion activity lie in the inner galactic regions (within $\sim$1~kpc from the black hole, BH; \citealt{downes1998,bryant1999}) and are thought to be responsible for the feeding of the BH and the halting of the SF through feedback mechanisms from the BH itself. 

The picture is complicated by the fact that the processes that drive inflows of gas towards the center are not univocally identified and seem to depend strongly on the scale considered. At kiloparsec scale, stellar bars are the most robust and efficient instabilities, constituting preferential directions of infall of the material towards the centre, possibly triggered by galaxy interactions. At the edges of the bars, shock fronts and dense dusty star forming regions are not unusual as a result of the sudden change of direction of the spiralling infalling material
(e.g., \citealt{saito2017}). Similarly, it is not uncommon to identify dynamically decoupled bars on different angular scales with overlapping dynamical resonances (\citealt{combes2013} and reference therein). In fact, on smaller scales, secondary nuclear bars seem to proceed from the primary bars (e.g., \citealt{shlosman2002}), and on $\sim$10-100pc scales, nuclear disks become unstable. On even smaller scales, the clumpiness of the interstellar medium (e.g., dynamical friction and viscous torques due to turbulence) starts to play a major role (e.g., \citealt{combes2012}).

In recent years, outflows of either ionised or molecular gas from the SMBHs have been observed and considered responsible for halting the infall and the SF in the inner galactic regions (e.g., \citealt{feruglio2010,cicone2014,harrison2016}). Fuelling of the SMBH and outflows seem to constitute the self-regulating combination of processes on the small scales responsible of the morphology and dynamics of the whole galaxy. 
The co-existence of such different processes, how the small scales dynamics influences the overall galaxy morphology, the timescale on which different processes happen, and if 
different evolutionary stages justify the different observed morphologies are still open questions.

These processes can now be investigated thanks to the unprecedented sensitivity and resolution of the Atacama Large Millimetre Array (ALMA), with the observation of CO molecule transitions in local Seyfert galaxies offering the possibility of directly witnessing the molecular gas fuelling towards the nucleus, thus allowing for the first time a thorough knowledge of the accretion process in the innermost regions of AGN (i.e., on pc scales from the nucleus; see, e.g., \citealt{garciaburillo2016}).
  
For a few nearby Seyfert galaxies, sub-mm/mm continuum and spectroscopic observations are starting to become available in the ALMA archive.
Recently, \citet{gruppioni2016} have collected the multi-wavelength photometric data for a subsample of 76 Seyfert galaxies from the IRAS 12-$\mu m$ galaxy sample (12MGS, \citealt{rush1993}) selected for having available Spitzer-IRS and literature mid- and far-IR spectroscopic data. The sample, though incomplete for statistical studies, being randomly selected from a nearly complete sample,  maintains a robust statistical significance and can be considered as representative of the local Seyfert galaxy population (as confirmed by a two-sided Kolmogorov-Smirnov test; see \citealt{gruppioni2016}). 
The photometric data available in the literature -- from the radio to X-ray bands -- allowed for this sample a detailed spectral energy distribution (SED) decomposition (see an example in Figure~\ref{fig:sed}),  that, including the contribution of stars, re-emission from dust in star forming regions and from a AGN dusty torus, provided precise measurement of the main properties of AGN and its host galaxy, like, e.g., the AGN bolometric luminosity, the star formation rate (SFR), the total IR luminosity, and the stellar mass. 

For one of these Seyfert galaxies, NGC~5135, ALMA data were available in the archive in two bands (both continuum and CO lines): band 6 (B6, $\sim$1.3 mm) and band 9 (B9, $\sim$500 $\mu$m). In this paper we present the ALMA data, and discuss possible interpretation of the resulting complex morphology and dynamical structure of this galaxy. To this purpose, we considered the available $\sim$100 pc resolution ALMA B6 and B9 observations of dust continuum emission, the spectroscopic maps of two CO transitions (tracer of molecular mass in star forming and nuclear regions), and of a CS transition (tracer of the dense star forming regions) and the outcome of the SED decomposition. 
NGC5135, thanks to its almost face-on appearance, the presence of a large scale bar, the bulge overdensity of stars, the past evidences of the presence of a Supernova Remnant (SNR) and a central AGN, 
is an excellent target to investigate the dynamics of inflows, outflows, SF and AGN feedback. 
\begin{figure}
 \includegraphics[width=\columnwidth]{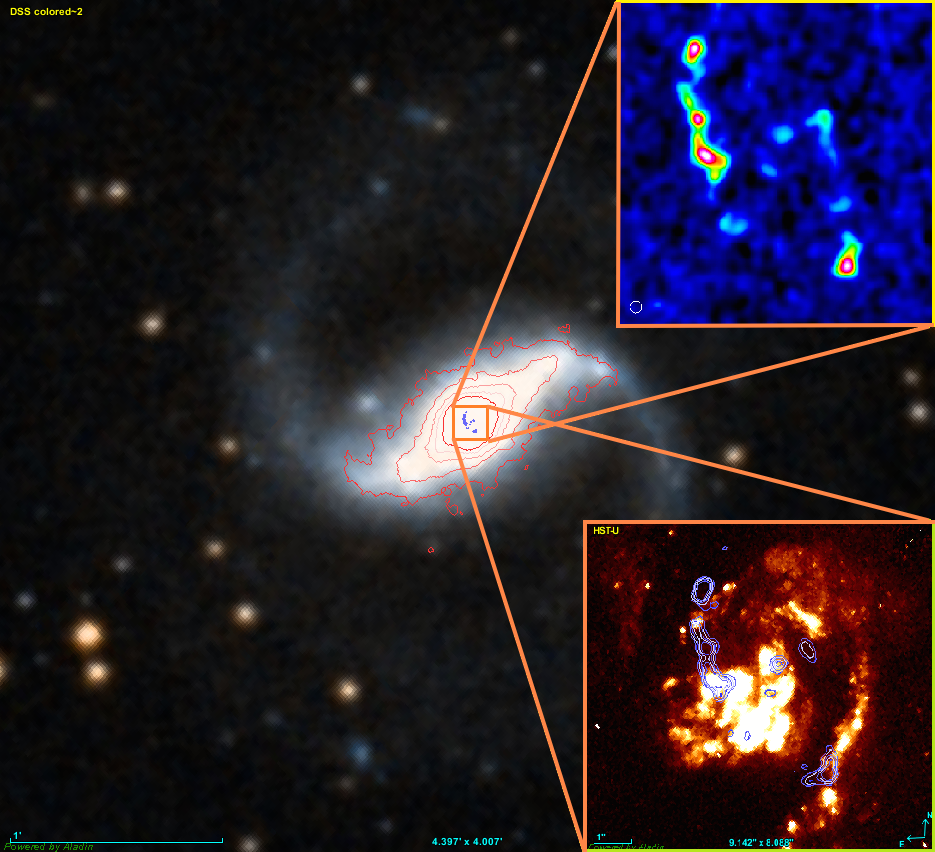}
 \caption{Optical DSS image, with overlaid contours from 2MASS (red) and ALMA B9 (blue, center). The top right insert shows a zoom of 
the ALMA B9 continuum image, while the bottom right one shows the ALMA contours overlaid to the HST $F606W$ image by \citet{gonzalezdelgado1998}.}
 \label{fig:sovrapposizione}
\end{figure}

In $\S 2$ we present the information on NGC~5135 available from the literature, and the SED decomposition and IR spectroscopy results. 
In $\S 3$ we present the ALMA archival data for this object and the data reduction/calibration that we have performed. In $\S 4$ we discuss the complex
gas morphology and kinematics in the inner regions of NGC~5135 and the possible responsible mechanisms. In $\S 5$ we summarise our results.

\noindent Throughout this paper, we use a Chabrier (2003) initial mass function (IMF) and we adopt a $\Lambda$CDM cosmology with $H_{\rm 0}$\,=\,71~km~s$^{-1}$\,Mpc$^{-1}$, $\Omega_{\rm m}$\,=\,0.27, and $\Omega_{\rm \Lambda}\,=\,0.73$. 

\section{NGC5135}
\begin{figure}
\centering
\includegraphics[width=7.5cm]{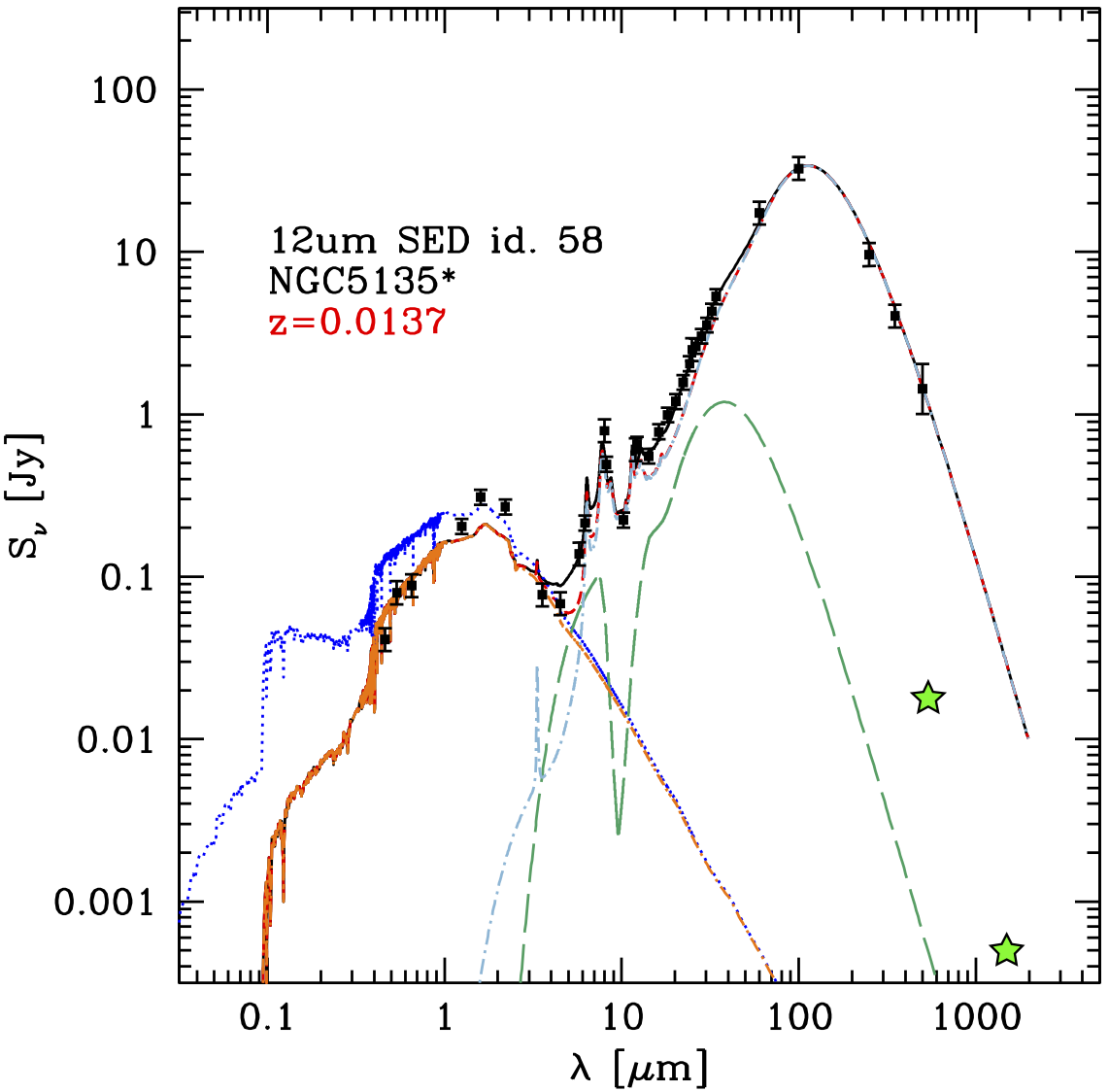}
\caption{Broad-band SED of NGC 5135 (data from the literature), as fitted by \citet{gruppioni2016} with a three component model -- extincted stars (red), AGN (green) and dust re-emission (pale blue) -- using the SED decomposition method described in \citet{berta2013}. The electric-blue emission at short wavelengths shows the expected unextincted stellar contribution. The green stars show the ALMA B6 and B9 data corresponding to the central region 
(AGN dusty torus?) obtained from our analysis.}
 \label{fig:sed}
\end{figure}
NGC~5135, a nearby galaxy at $z$=$0.013693$ (implying a scale of $\sim$281~pc/arcsec), was first selected by \citet{phillips1983} as part of a sample of $23$ nearby high-excitation emission-line galaxies, from the original catalogue of 719 bright galaxies by \citet{sandage1978}. 
It was then classified as a Seyfert 2 by \citet{veroncetty2006} based on the [O~III]$\lambda$5007/H$_{\beta}$ and [N~II]$\lambda$6584/H$_{\alpha}$ line ratios (BPT diagram: \citealt{baldwin1981}).
From its total IR luminosity (i.e., $L_{\rm IR}$$=$$L$[8--1000 $\mu$m] integrated), log($L_{\rm IR}$/ L$_{\odot}$)$=$11.17, NGC 5135 was then classified as a Luminous Infrared Galaxy (LIRG) by \citet{sanders2003}.
Its DSS image is shown in Figure~\ref{fig:sovrapposizione}, with overplotted the ALMA B9 continuum contours: the ALMA map is zoomed in the top right corner of the figure, while
the zoom at the bottom right corner shows the same ALMA B9 contours on top of the HST (F606W) image (\citealt{gonzalezdelgado1998}), while the broad-band SED and the result of the decomposition 
performed by \citet{gruppioni2016} are shown in Figure~\ref{fig:sed}.
From the broad-band (from UV to mm) SED fitting and decomposition, \citet{gruppioni2016} have estimated the main physical parameters of NGC~5135, such as the stellar 
mass, $M_{*}$$=$5.16$\times$10$^{10}$ M$_{\odot}$, the total IR luminosity, $L_{\rm IR}$$=$10$^{11.23\pm0.03}$ L$_{\odot}$, the star formation rate, $SFR$$=$(15.6$\pm$1.9)~M$_{\odot}$/yr, 
and the AGN bolometric luminosity, $L_{\rm BOL}^{\rm AGN}$$=$10$^{44.3\pm0.1}$~erg/s. The values derived from the SED fitting correlate well with the luminosities of the
mid-IR lines tracing SF (e.g., PAHs, [Ne~II], [Si~III]) and AGN activity (e.g., [Ne~V], [O~IV]), measured by \citet{tommasin2008,tommasin2010} from high resolution {\em Spitzer}-IRS spectra. 
 
Nuclear studies of NGC~5135 showed a complex morphology and the presence of numerous coexisting structures in the region within 0.7 kpc from the AGN, including a nuclear bar (e.g., \citealt{mulchaey1997}).
In order to describe the central engine, \citet{bedregal2009} observed NGC 5135 with VLT SINFONI in [Si~VI]$\lambda \:1.96\: \mu m$ line, and H-band ($1.45-1.85\: \mu m$) and K-band ($1.95-2.45\: \mu m$) continuum.  The [Si~VI] emission, that requires an ionisation energy of 166.7eV, is considered a good tracer of the AGN activity in the areas surrounding the nucleus, and it is mainly produced by gas excited just outside the broad line regions (\citealt{bedregal2011}).

Our target was also observed by \citet{levenson2004} in the X-rays with Chandra (0.4-8 keV), and the AGN emission component was separated from the stellar one. Two different components were identified in the central X-ray sources: a northern component associated to the AGN, and a southern associated to a star forming region (hereafter called SFr). The SFr found in the central region seems to contribute to the high obscuration of the AGN, producing an hydrogen column density $N_{H}>10^{24}\:cm^{-2}$. 
The SF regions were previously observed also by HST (\citealt{gonzalezdelgado1998, alonsoherrero2006}) and by VLT SINFONI (in Br$\gamma$ $\lambda \:2.17 \mu$m;  \citealt{bedregal2009}). Finally, a Supernova remnant (SNR) and a massive outflow (likely supernova driven) located in the Southern part of the nuclear region, were found by using VLT SINFONI observations of  [Fe~II] $\lambda \:1.64\: \mu m$ as a tracer (\citealt{colina2012}). 

At radio wavelengths, \citet{ulvestad1989} found an intense emission at 6 and 20 cm in VLA-observations, with total fluxes of 163.2 mJy and 58.8 mJy respectively. The complexity of the radio emission, that presented a diffuse faint North emission with a bright source in the Southern region in both bands, allowed the authors to classify NGC 5135 as an ''ambiguous'' radio source.

In order to complete this scenario, we have analysed ALMA archival data, adding the cold dust and molecular gas view to the great wealth of available information about NGC~5135. 

\section{The ALMA (archival) data}
\label{alma}

\begin{figure*}
\includegraphics[width=19cm,height=17cm]{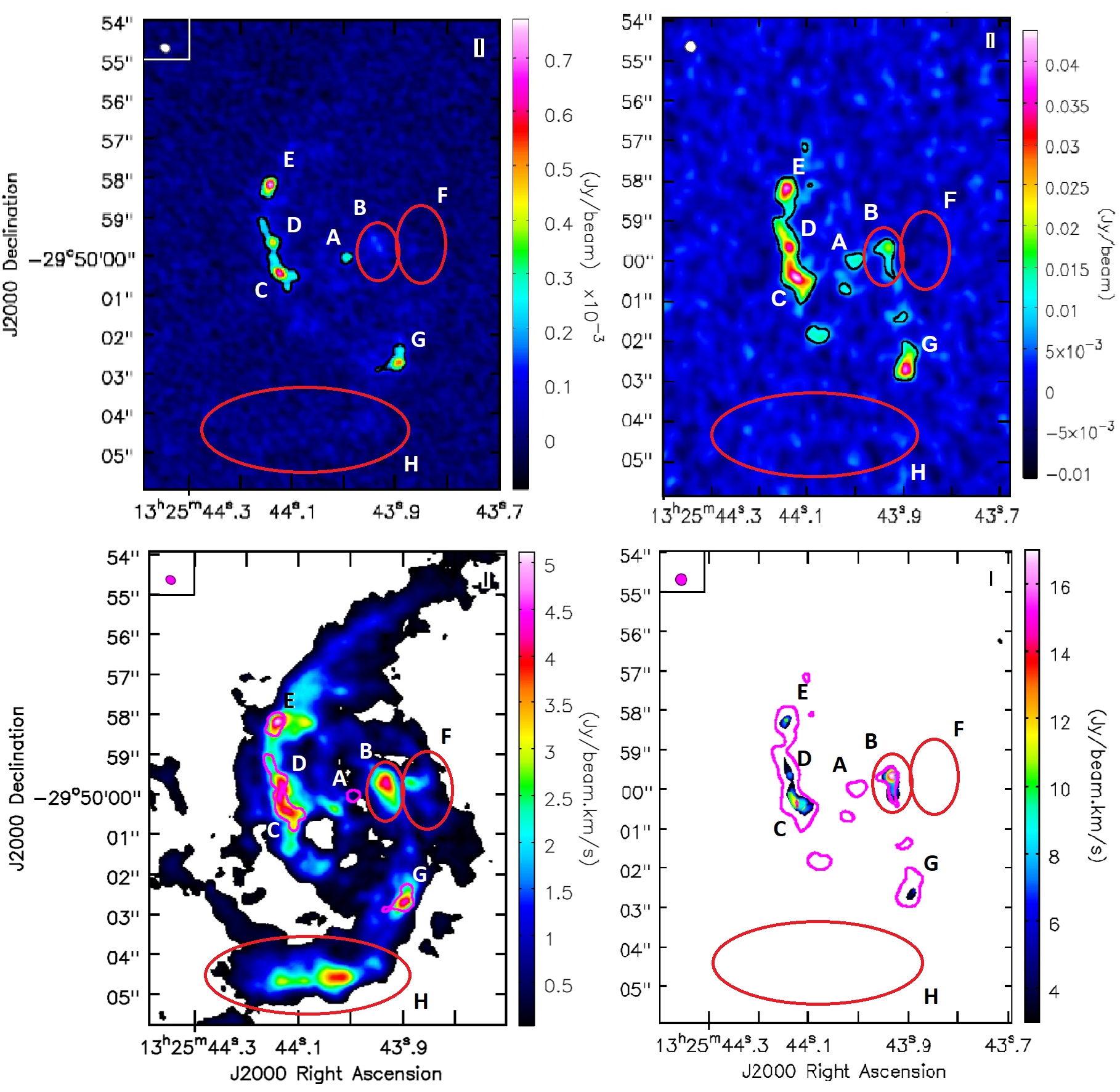}
\caption{ALMA images of NGC~5135 in B6 ($left$) and B9 ($right$). Continuum emission (in mJy/beam, $top$) and moment 0 (i.e., gas distribution, in units of Jy km s$^{-1}$ beam$^{-1}$, $bottom$) 
of CO(2--1) ($left$) and CO(6--5) ($right$). To avoid the noise altering the color scale, we have imposed a 3$\sigma$ threshold to the CO(2--1) and a 4$\sigma$ threshold to the CO(6--5) moment 0 maps. We overplot the 4$\sigma$ contour level in the same band to each continuum map (shown in black), and to the moment 0 maps (shown in magenta). 
The colour bars on the right show the scale of the continuum and line intensities.
The letters identify the different regions discussed in the text. At the $top$ $left$ corner of each image we show the corresponding beam as a coloured ellipsis.}
\label{fig:ALMAimages}
\end{figure*}

\begin{figure*}
\includegraphics[width=19cm,height=17cm]{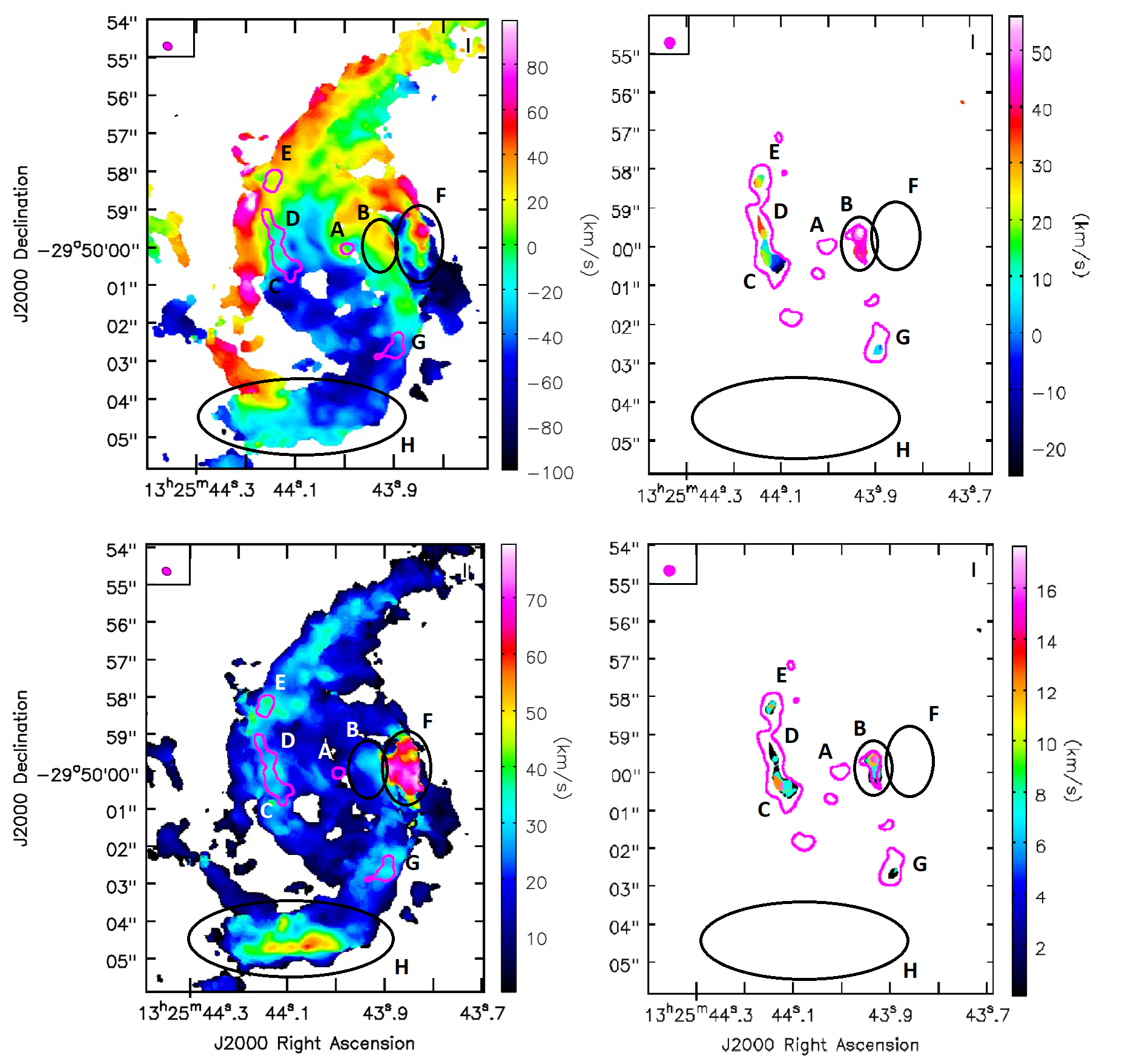}
\caption{Moment 1 (velocity field; $top$) and 2 (velocity dispersion; $bottom$) of CO(2--1), detected in B6 ($left$), and CO(6--5), detected in B9 ($right$). 
The beam sizes, thresholds, contours, letters and marked regions are the same as in Figure~\ref{fig:ALMAimages}.}
\label{fig:ALMAmom}
\end{figure*}

ALMA B6 observations were obtained in Cycle 3 (project 2013.1.00243.S, PI: Colina) in two different configurations of the 12-m antennas array. 
The ``compact'' configuration was obtained with 38 antennas distributed over the baseline range between 15.0 and 348.5 m, while the
``extended'' one with 44 antennas over the baseline range 15 -- 1600 m, reaching resolutions of 0.16 arcsec at the sky frequency. 
The combination of the two configurations allows to cover angular scales in the range 0.1 -- 12.32 arcsec. 
These scales correspond to spatial scales of 28.9 pc -- 3.56 kpc at the redshift of NGC~5135. 
Two 1875 MHz spectral windows, centred respectively at the sky frequencies of 227.4 GHz, and 241.6 GHz, cover the expected frequency of CO(2--1) and CS(5--4) molecular line emissions with a channel width of 488 kHz (corresponding to 0.42~km$s^{-1}$ at the observing frequency). 
A third 2-GHz wide spectral window centred at 228.8 GHz, with a channel width of 15.625 MHz, was dedicated to improve the continuum characterisation. 
The data available in the archive were calibrated using the ALMA calibration pipeline using CASA version 4.2.2. 
The calibrated data-sets of the two configurations were concatenated and combined with CASA 4.7.1. 
The achieved continuum beam size of the concatenated image is 0.25$^{\prime \prime}\times$ 0.20$^{\prime \prime}$, while the 1$\sigma$ RMS is 1.9$\times$ $10^{-5}$ Jy beam$^{-1}$. 
The maximum recoverable scale allowed by the array configuration is $\simeq$12 arcsec.
Finally, the CASA standard cleaning was performed on the combined image, considering \textit{briggs - $robust=0.5$} weighting.  

ALMA B9 observations (cycle 3; project 2013.1.00524.S, PI: N. Lu) were taken on June 2$^{nd}$ 2015 with an array composed by 39 12-m antennas included in a range of baselines between 20.5 and 886 m. Four spectral windows were centred on the sky frequencies 678.2GHz, 680.2GHz, 683.7GHz and 682.0GHz respectively. At these frequencies, the angular scales observed are in the range 0.06 -- 2.68 arcsec, corresponding to 17.33 pc -- 774.38 pc at the source redshift. The expected sky frequency of CO(6--5) transition falls in the fourth spectral window.\\
Manual calibration was performed for the archival data. The phase calibrator ($J1316-3338$) is faint at these observing frequencies ($\sim$ 0.4 Jy).
With respect to the script available in the archive, we improved the calibration by flagging four mis-behaving antennas and updated the flux density model for the flux density calibrator 
by a factor 0.84 (to the value of 1.39 Jy), 
exploiting an observation of the calibrator coeval to those of the target (not yet available in the calibration catalogue at the time of the quality assessment analysis). 
The beam of the B9 continuum image is 0.30$^{\prime \prime}$v$\times$0.29$^{\prime \prime}$, while the reached 1$\sigma$ RMS is 2.11$\times$ $10^{-3}$ Jy beam$^{-1}$. 
The maximum recoverable scale for the B9 configuration is $\simeq$2.68 arcsec.

We must stress that, given the smaller size of the maximum recoverable scale in our B9 configuration, any direct comparison between the fluxes measured 
in the two bands will not be
possible, since B9 might miss (i.e., filter out) diffuse fluxes -- if any -- on scales longer than the maximum recoverable ones.

\begin{figure}
\centering
\includegraphics[width=85mm]{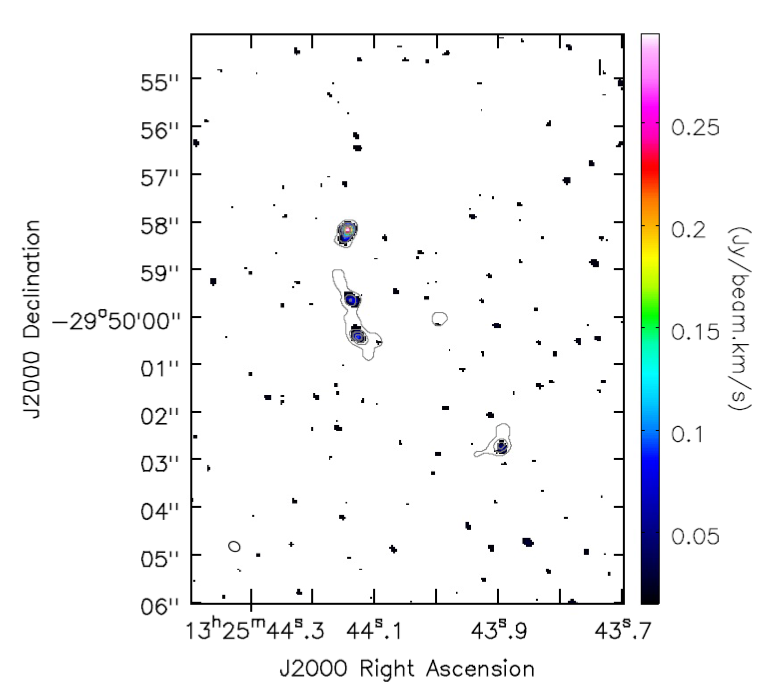}
\caption{Map of the CS(5--4) line. The contour levels show the CS emission at $>$4$\sigma$, while the colour bar shows the line intensity in Jy km s$^{-1}$ beam$^{-1} $.}
\label{fig:CS}
\end{figure}

The continuum images resulting from our ALMA data analysis are shown in the top panels of Figure~\ref{fig:ALMAimages} (B6 $left$ and B9 $right$). 
The 0$^{th}$ moment (i.e., intensity map in Jy km s$^{-1}$ beam$^{-1}$) of the CO(2--1) ($left$) and CO(6-5) ($right$) are shown in
the $bottom$ panels of the same figure.
The 1$\sigma$ RMS obtained for the moment 0 maps is 1.1$\times$  $10^{-3}$ and 1.5$\times$ $10^{-2}$ Jy km s$^{-1}$ beam$^{-1}$, respectively for CO(2--1) and CO(6--5).
In Figure~\ref{fig:ALMAmom} we show the velocity field ($top$) and the velocity dispersion ($bottom$) for the CO(2--1) and (6--5) transitions (on the $left$ and $right$ respectively), while in 
Figure~\ref{fig:CS} we show the intensity map of CS (5--4), for which we have obtained an 1$\sigma$ RMS of  1.5$\times$ $10^{-4}$ Jy km s$^{-1}$ beam$^{-1}$.
These images show very clearly the complex gas and dust distribution in this source, with hints of a central bar structure connecting two potential spiral arms. 

\begin{figure*}
\begin{center}
\includegraphics[width=\textwidth]{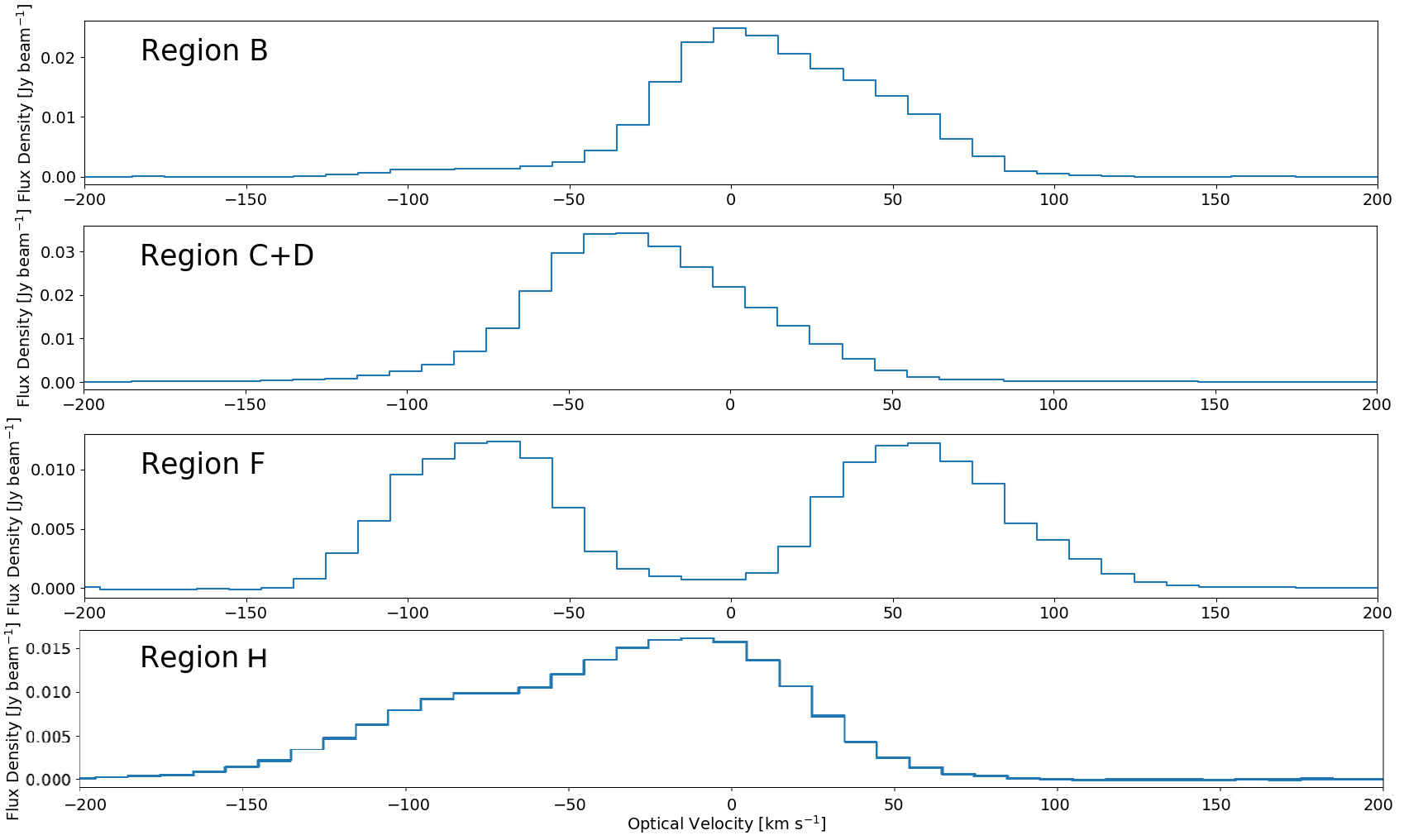}
\caption{CO(2--1) spectra of the regions ``{\tt B}'', ``{\tt C$+$D}'', ``{\tt F}'' and ``{\tt H}'' (from $top$ to $bottom$). The locations of these regions are shown in Figures~\ref{fig:ALMAimages} and \ref{fig:ALMAmom}.}
\label{fig:mom0_spec}
\end{center}
\end{figure*}

Several regions deserving particular attention
and interpretation are identified by combining the information derived from the continuum and the CO moment maps.
In order to select the boundaries for these regions (labelled with letters and if needed encircled by ellipses in Figures~\ref{fig:ALMAimages} 
and \ref{fig:ALMAmom}), and measure their fluxes, we have used the CASA-viewer. We have considered
the lower resolution band (B9 in our case, due to the antennae configuration), then we have
selected the extraction region as to include all the flux in that band in the continuum image,
applying the same extraction region to both bands.
Note that, although the flux across the three SF regions (``{\tt C}'', ``{\tt D}'', ``{\tt E}'') extends for about 3 arcsec in
declination, we have tried to measure the flux connected to the single knots separately by fitting the three regions singularly.
The observed CO(2--1) spectra for the most relevant emitting regions,
are shown in Figure~\ref{fig:mom0_spec}, and are described in detail in the following paragraph.

\begin{itemize}

\item Region ``{\tt A}'' (centred at RA(J2000) $+13:25:43.99$ DEC(J2000) $-29:50:00.02$, as found from the B6 continuum image; see Section~\ref{3DBarolo}) includes the AGN position, as identified in VLT-SINFONI [Si IV]$\lambda\:1.96\:\mu m$ and Chandra X-ray maps (RA(J2000) $+13:25:44.0$ DEC(J2000) $-29:50:01$; \citealt{bedregal2009,levenson2004}). It is clearly visible in both the continuum maps, implying the presence of cold and dense dust in the very central regions, but it is not clearly detected in the observed spectral line transitions. Similarly, \citet{Schinnerer2000} did not detect the nucleus of NGC~1068 in CO, although very recently the NGC~1068 torus and multiple knots of SF have been resolved in CO with ALMA at much higher resolution (4.2 pc; \citealt{garciaburillo2016}).
Similarly to what happens in NGC~1068, where the AGN is part of a circum-nuclear disk that could be resolved only with very high resolution observation (\citealt{garciaburillo2016}), 
the flux from region ``{\tt A}'' can be contaminated by SF and only partially due to the AGN dusty torus. However, at odds with the dust distribution in NGC~1068, the continuum emission from region ``{\tt A}'' in NGC~5135 appears unresolved at the resolution of the B6 data, but already isolated (i.e., not embedded in a circum-nuclear ring, which is expected to extend out to larger distance). 
We stress that the circum-nuclear disk observed by \citet{garciaburillo2014} in NGC~1068 at a resolution of 60--70~pc was successively resolved into both AGN torus and SF knots, once observed at much higher resolution (4.2~pc; \citealt{garciaburillo2016}). Therefore, given fact that the B9 
emission in region ``{\tt A}'' is significantly larger than the beam (see the 4$\sigma$ contours in Figure~\ref{fig:ALMAimages}), 
and given the resolution of these data, we are not able to separate out SF knots (if any) and AGN.
In any case, since it is not possible to estimate the continuum emission from the AGN torus without SF contribution unless the AGN torus is clearly resolved, 
we can consider the flux measured from region ``{\tt A}'' as an upper limit to the cold dust emission component from the torus.   
The ALMA B6 (1.3 mm) and B9 (450 $\mu$m) continuum measurements corresponding to the central region (e.g., ``{\tt A}'') are shown as green stars in Figure~\ref{fig:sed}. 
Our ALMA measurements seem to show much higher cold dust contribution from the nucleus than expected from torus only models. 
ALMA higher resolution measurements would be necessary for this galaxy, in order to be sure to properly model and interpret the emission from the AGN (i.e., without any SF contamination), 
so far totally unconstrained by data in the Rayleigh Jeans tail. Only by resolving the torus emission itself we can derive its real contribution, although if the flux measured in the central 
region will come out to be mostly due to the AGN, this result will challenge the current torus models, implying the presence of a colder dust component.
However, as previously noticed, the B9 continuum measurement can just be taken as an indication and is not directly comparable with the B6 one, since the smaller recoverable scales in the B9 configuration filter out signals at larger scales, if any. 
\item Regions ``{\tt B}'' and ``{\tt C}'' correspond to the bar edges. The sudden change of the in-falling direction in this region seems to produce a dense and hot environment where SF is active. 
The CO(2--1) spectrum of region ``{\tt B}'' peaks at the galaxy velocity (i.e., 0 km s$^{-1}$ in Figure~\ref{fig:mom0_spec}), but shows a positive velocity tail (possibly a secondary peak at about 40-50 km s$^{-1}$), while the spectrum of the  ``{\tt C$+$D}'' regions peaks at negative velocities (e.g., $-$30/$-$40 km s$^{-1}$).
Regions ``{\tt C}'', ``{\tt D}'' and ``{\tt E}'' had been already identified as SF loci by \citealt{bedregal2009}: this is confirmed now by the presence of strong CO emission.  
Region ``{\tt E}'' is also the dominant region in the CS(5--4) brightness distribution (flux density $\sim$0.29$\pm$0.03 Jy/beam, see fig. \ref{fig:CS}), thus likely to be denser than the others. A detailed study of the main properties, physical parameters and SF rates in these regions will be presented and discussed in a further paper from our team, where a detailed chemical analysis will also be performed.
\item Region ``{\tt G}'' corresponds to the SNR detected by HST and VLT SINFONI observations, and clearly visible here in all the observed emissions. Observation with VLA at 6 cm (\citealt{ulvestad1989}) also shows a strong emission at this location, suggesting strong synchrotron emission from SNR. 
\item Two regions of high velocity dispersion, namely regions ``{\tt F}'' and ``{\tt H}'', were not detected in any previous observations:
\subitem{\bf --} Region ``{\tt F}'' is aligned along the bar, but seems to lie outside its edge. It is characterised by a double peaked spectrum (see Figure~\ref{fig:mom0_spec}) with two opposite velocity components ($\sim$$\pm$70 km s$^{-1}$). It is not detected in continuum emission.
\subitem{\bf --} Region  ``{\tt H}'' lies along the southern arm and appears only in the CO(2--1) maps, not in continuum or CO(6--5), showing a high velocity dispersion 
and a secondary peak in the spectrum corresponding to large negative velocity ($\sim$ $-$70/$-$80 km s$^{-1}$; see Figure~\ref{fig:mom0_spec}).
\end{itemize}

\section{{\tt $^{\rm 3D}$BAROLO} kinematic modelling}
\label{3DBarolo}
In order to investigate the kinematics n the innermost regions of NGC~5135, we used the {\tt$^{\rm 3D}$BAROLO} 
software to fit our ALMA data cubes. {\tt$^{\rm 3D}$BAROLO} (3D-Based Analysis of Rotating Object via Line Observations) is a code developed by \citet{diteodoro2015} which derives rotation curves of galaxies from emission line data  cubes by fitting a 3D tilted-ring model. 
The tilted-ring model is based on two assumptions: the material which is responsible for the emission lines is contained into a thin disc, and the kinematics is dominated by rotation. 
In this model, the galaxy disc is divided into a series of concentric rings characterised by seven free parameters (three of which describe the morphology and four the kinematics): the kinematical center ($x_{0},y_{0}$), the inclination with respect to the observer ($i$) and the position angle of the major axis ($\phi$) define the characteristics of the rings projected on the sky plane; the velocity dispersion ($\sigma$), the rotational velocity ($V_{rot}$), the systemic velocity ($V_{sys}$) and the radial velocity ($V_{rad}$) of the gas, define the kinematic characteristics of the model.
In the {\tt$^{\rm 3D}$BAROLO} version used in this work (v1.3), the three velocity components are combined to define the observed velocity along the line of sight ($V_{LoS}$), mapped by 
moment 1. The derived velocity map can be written in terms of a generic harmonic expansion (e.g., \citealt{schoenmakers1997, swaters1999}) 
that we can write as (\citealt{roelofs1995} for details):

\begin{equation}\label{VLOS}
V_{LoS}(x,y)=V_{sys}-sin\: i[V_{rot}(R)cos\: \varphi -V_{rad}(R)sin\: \varphi]
\end{equation}

\noindent
where $\varphi$ is the position angle of the major axis on the receding half of the
galaxy, taken anticlockwise from the North direction on the sky.
The code finds the best-fitting values by minimising the difference between the data-cube and the kinematic model smoothed to the same resolution of the data.
{\tt$^{\rm 3D}$BAROLO} provides three methods for calculating the residuals by comparing the model and the data for every pixel in the 3D space of the data-cube: in our case, we 
have chosen a kind of $\chi^{2}$, described as follows. 
If M is the flux associated with a model-pixel and D is the flux of a data-pixel (the flux of the RMS of the data-cube is associated to D in case it is a blank pixel; 
see \citealt{diteodoro2015} for details), then {\tt$^{\rm 3D}$BAROLO} minimises the quantity:

\begin{equation}
\Delta r= \frac{(M-D)^{2}}{\sqrt{D}} ,
\end{equation}


\noindent
A further parameter to be specified is the column density of the gas. For this parameter, two different normalisations are implemented in {\tt$^{\rm 3D}$BAROLO}: pixel-by-pixel (i.e., \textsf{local}) and azimuthally averaged (i.e., \textsf{azim}). In the former case, the model assumes equal value for the integral of each spatial pixel along the spectral dimension in the model and in the observations. This normalisation allows us to consider non-axisymmetric density models and avoids that regions with anomalous gas distribution, like e.g. clumpy emission or holes, affect the global fit. The normalisation value of the second model is instead set to the azimuthally averaged flux in each ring. 
Although the \textsf{azim} normalisation cannot reproduce the details of the galaxy morphology (which are instead reproduced on a pixel-by-pixel basis by the \textsf{local} normalisation), 
it provides useful information about some of the model parameters (e.g., the inclination angle $i$) and is important to validate the {\tt local} model solutions. 
In fact, since the observed gas distribution is non axisymmetric, it is not obvious that the solutions found by the two different normalisation models coincide, 
in case the result is not the correct one: to confirm that the kinematic model correctly reproduces the observed global motion of the gas, the {\tt local} model solutions 
must be compatible with those found by the {\tt azim} one.

\begin{figure}
 \includegraphics[width=\columnwidth]{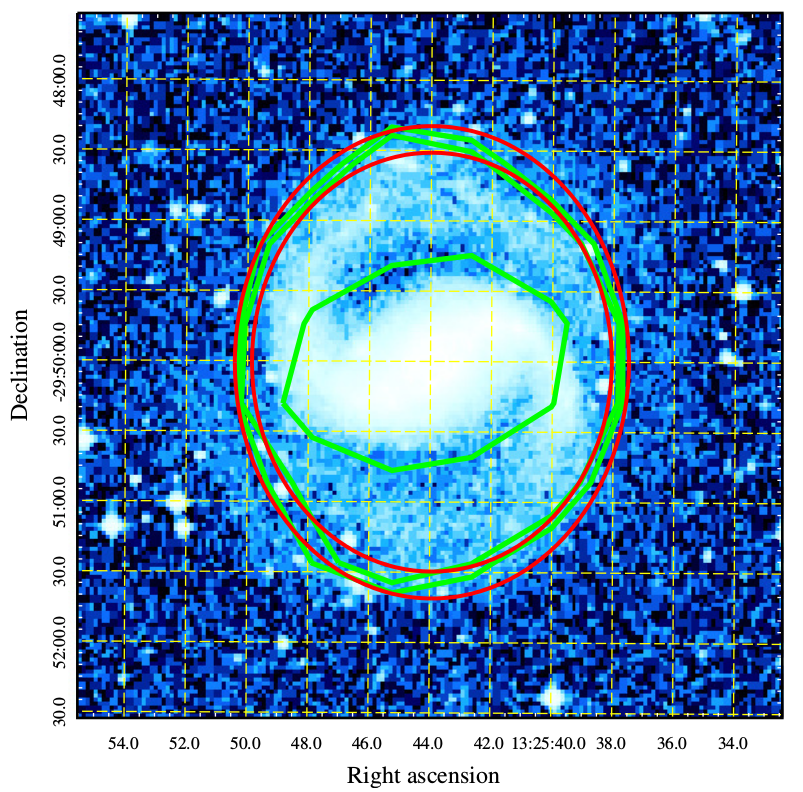}
 \caption{Optical Palomar image of NGC5135 in histogram colour scale. Smoothed (with \textsf{smoothpar}$=$15) galaxy's contours are overplotted in green. 
 The red curves are the internal and external ellipses matching the outer isophotal (green) contour.}
 \label{fig:elliptic}
\end{figure}

\noindent
Some of the model parameters have been fixed in order to reduce the computation-time and the degeneracy. 
For some of them the values have been fixed based on assumptions about the gas distribution (e.g., we have fixed the thickness of the disk by assuming a thin disk distribution), 
while other parameters were measured from the data, then fixed after checking their validity. 
For instance, the kinematic center - derived by setting a circular region around the AGN and assuming the emission peak of the B6 continuum coincident with the position of the AGN - 
was initially let free to vary. Once verified that {\tt $^{\rm 3D}$BAROLO} was finding the same kinematic position (within errors), we have fixed its value. 
Other two parameters were fixed in our model, in addition to the galaxy centre: the position angle ($\phi$) and the inclination of the galaxy disc ($i$). 
By using the code \textsf{kpvslice}, an extension of the \textsf{Karma} software (\citealt{gooch1996}), we found $\phi \sim10^{+5}_{-10}$ deg as the best approximation 
of the major (projected) axis orientation.
The inclination ($i$) of the disc is estimated by using the DSS image of NGC~5135 (see Figure~\ref{fig:elliptic}) and matching the isophotal contours of the outer disc
with an ellipsis. The inclination of the major axis of the outer ellipsis defines the inclination of the disk. 
This method provided an estimate of the inclination angle $i$=$57\pm 2$ deg, with the associated error derived by considering the average difference between the two red ellipses in 
Figure~\ref{fig:elliptic}.

\subsection{Results: Kinematic models for molecular gas of NGC~5135}

\begin{figure*}
\begin{center}
\includegraphics[width=170mm]{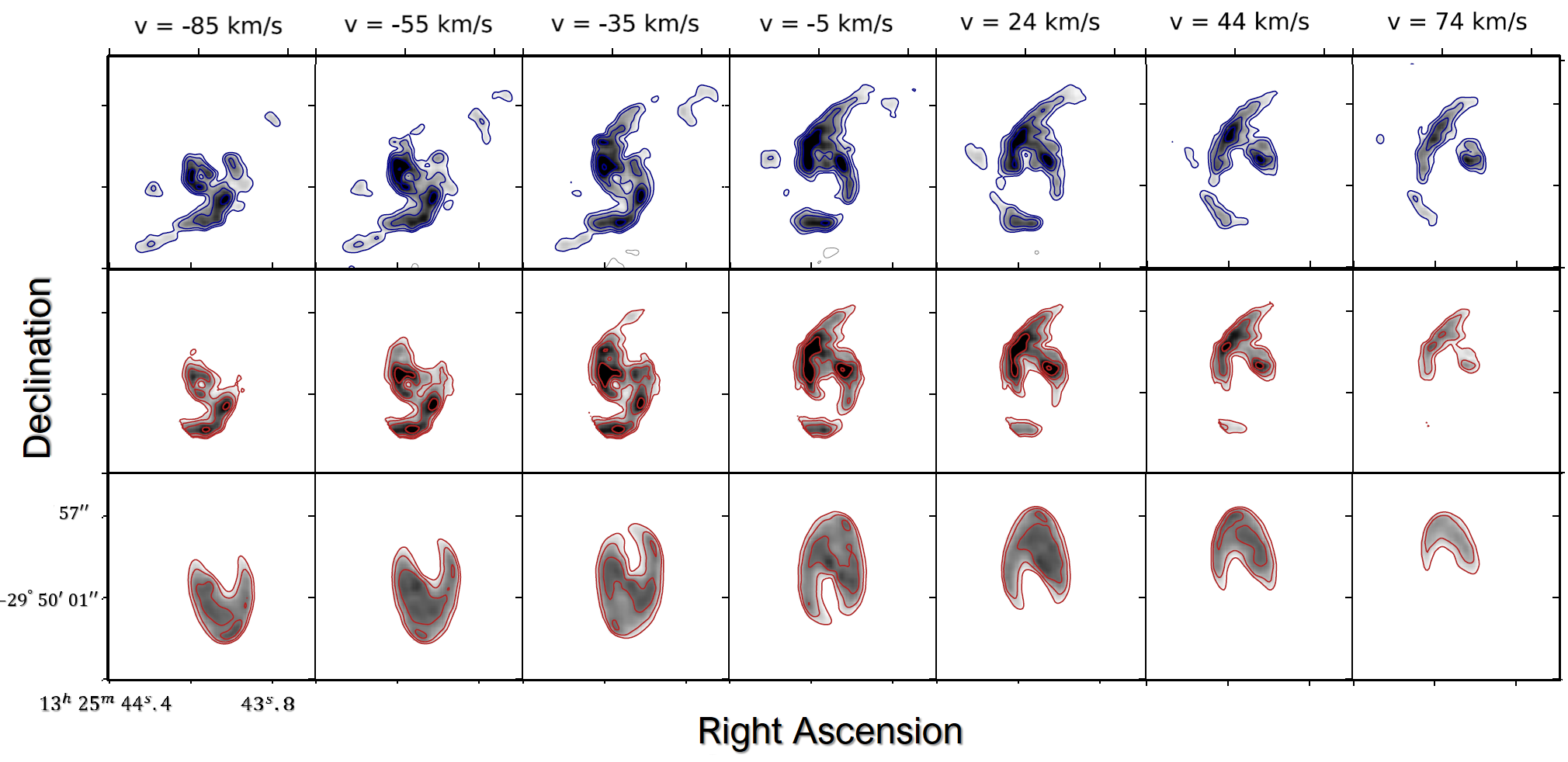}
\caption{Selected CO(2--1) channels from the ALMA B6 data cube (in blue, top row), compared with the {\tt local} normalisation mode (in red, middle row) and the {\tt azim} normalisation mode (red, bottom row). All models are obtained considering a threshold of $4\sigma$.}
\label{fig:modelli}
\end{center}
\end{figure*}
In Figure~\ref{fig:modelli} we show the results of our fit to the ALMA CO(2--1) data-cube (up to an angular size of 6 arcsec, corresponding to $\sim$1.7 kpc from the galaxy centre) with {\tt$^{\rm 3D}$BAROLO}, in seven representative velocity channels (out of 52) of spectral resolution 10 km s$^{-1}$. 
The velocity of the galaxy is centred on the CO(2--1) line emission peak at 227.43 GHz, derived from the continuum B6-concatenated observation at the central AGN position.
In the $top$ row of the figure we show the data velocity field, while in the $central$ and $bottom$ rows we plot the results of the {\tt local} and {\tt azim} model
respectively, in the same velocity channels. Again, we stress how the comparison between the {\tt azim} and the {\tt local} model is a probe of the goodness of the {\tt local} fit, 
since the {\tt azim} model reproduces the ``global'' behaviour of the data and contains the same amount of gas as the {\tt local} model. 
The region showing two opposite velocities (i.e., $v$$\simeq$$\pm$70 km s$^{-1}$, region ``{\tt F}''), cannot be reproduced by the models, while 
the other high velocity dispersion region located in the southern spiral arm (region ``{\tt H}'') can be only partially reproduced, with a tail at $\lsimeq$$-$40 km s$^{-1}$ not fitted by the model.

\begin{figure*}
\centering
\fbox{\includegraphics[width=\textwidth]{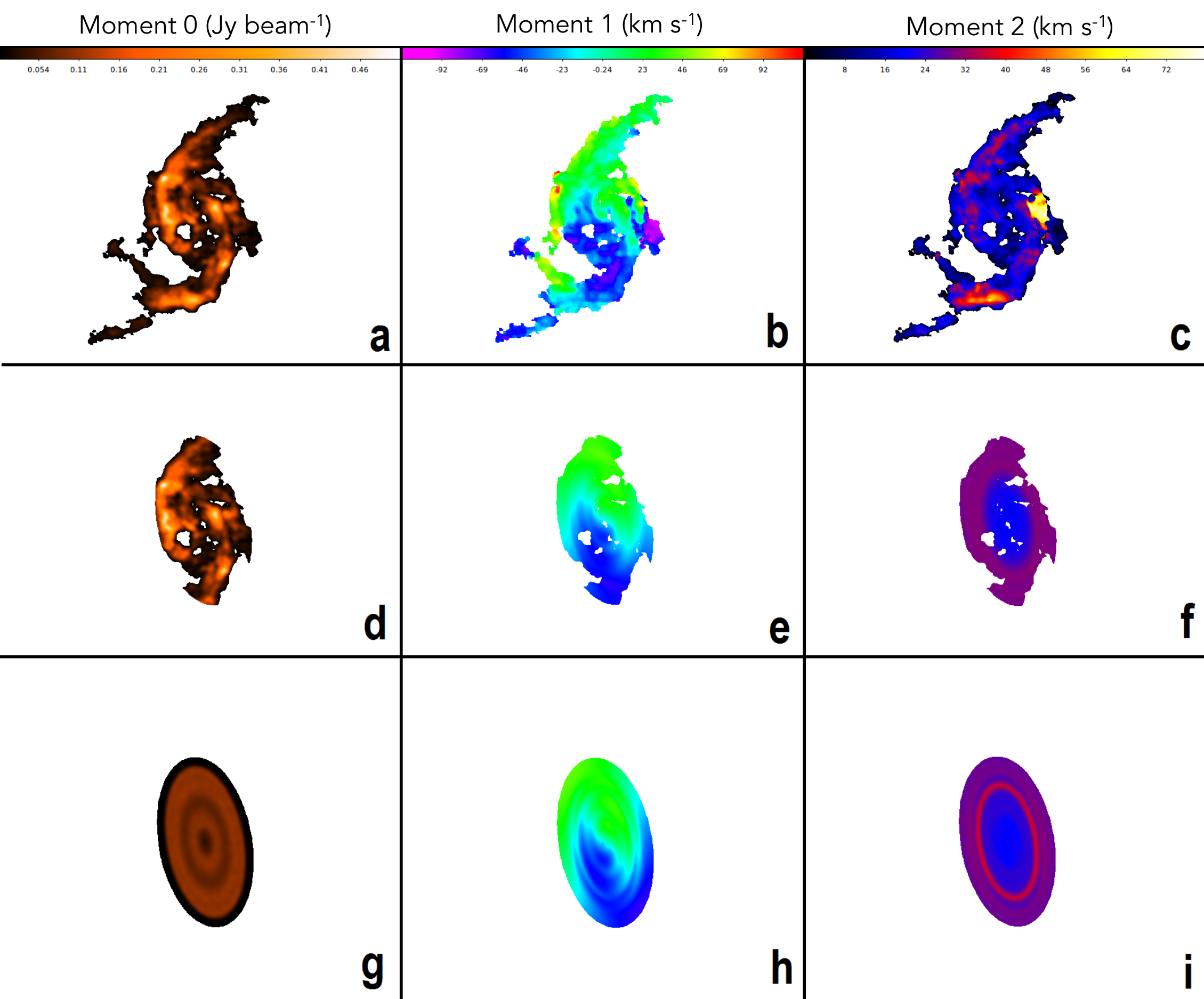}}
\caption{Molecular gas distribution (i.e., CO(2--1)) in the centre of NGC~5135 and the corresponding models obtained with {\tt$^{\rm 3D}$BAROLO}. 
Panels \textsf{(a)}, \textsf{(b)} and \textsf{(c)} show the $0^{th}$ (map), $1^{st}$ (velocity field) and $2^{nd}$ (velocity dispersion) moments respectively, obtained from our CO(2--1) B6-concatenated data 
cube. Panels \textsf{(d)}, \textsf{(e)} and \textsf{(f)} are the $0^{th}$, the $1^{st}$ and 2$^{nd}$ moments of the \textsf{local} normalisation {\tt $^{\rm 3D}$BAROLO} model, while
panels \textsf{(g)}, \textsf{(h)} and \textsf{(i)} are the same moments for the \textsf{azim} normalisation model. 
All moments are extracted at a threshold of $4\sigma$. Units are in Jy/beam for the CO(2--1) emission ($left$ column) and km s$^{-1}$ for the velocity field and velocity dispersion
($central$ and $right$ columns).
}
\label{fig:momentiTOT}
\end{figure*}

The comparison of our molecular gas distribution with the {\tt $^{\rm 3D}$BAROLO} model expectations gives us several important information about the gas structure and kinematics in the inner regions
of NGC~5135. 
In Figure~\ref{fig:momentiTOT} we show the three moments (0$^{th}$, 1$^{st}$ and 2$^{nd}$) of the CO(2--1) molecular gas distribution in the centre of NGC~5135 from the data cube ($top$ row), 
and from the \textsf{local} ($central$ row) and \textsf{azim} ($bottom$ row) models obtained with {\tt $^{\rm 3D}$BAROLO}.
Note that we have limited our moment analysis to the central regions of the galaxy, within a disc of radius $R_{max}$ $=$4$\arcsec$ ($\sim$1.167 kpc), 
because, as the ring radius increases, the CO(2--1) distribution strongly departs from axisymmetry, thus {\tt $^{\rm 3D}$BAROLO} cannot find acceptable solutions.
Both models reproduce reasonably well the observed velocity field (moment 1) of the CO(2--1), shown in panel \textsf{(b)} of Figure~\ref{fig:momentiTOT} (models in panels \textsf{(e), (h)}), 
while only the {\tt local} model provides a good fit to the observed gas distribution (see panel \textsf{(d)} versus \textsf{(a)}). 
In fact, since the \textsf{azim} model averages the brightness distribution ring-by-ring, 
it is not able (by definition) to reproduce the pixel-by-pixel asymmetry of the data (see panel \textsf{(g)} versus \textsf{(a)}).
The complex structure of the velocity dispersion map (expressed as line width) observed in the data, showing zones of very large velocity dispersion, not expected for a rotation 
disk with super-imposed radial motions, is not reproduced by any of the models. 
In particular, in panel \textsf{(c)} we clearly note the two zones of very large velocity dispersion already identified in Section~\ref{alma}: 
one at the edge of the potential bar, showing two opposite velocities corresponding to the same spatial position
(i.e., the bright yellow spot in panel \textsf{(c)} of Figure~\ref{fig:momentiTOT}, corresponding to region ``{\tt F}''), and the other in the southern arm of the spiral structure 
(i.e., the orange region in the figure, corresponding to region ``{\tt H}''). 
Being the model axisymmetric, it cannot reproduce these local regions of high velocity dispersion in the data, but
finds an averaged solution on the entire ring. This happens with both the {\tt azim} and the {\tt local} model.
We note however that {\tt $^{\rm 3D}$BAROLO} is nonetheless able to identify a gradient, in fact the larger values of the velocity dispersion in outer rings of panels \textsf{(f), (i)}, 
are due to the presence of region ``{\tt F}'' and are not intrinsic dispersions: without the high dispersion regions, the model would reproduce reasonably well the average observed 
values within each ring. 
Finally, we note that the presence of the nuclear bar at the centre of NGC~5135 introduces motions in the innermost gas that the {\tt$^{3D}$BAROLO} software is unable to reconstruct perfectly. 
Therefore, we cannot consider our best-fit model as fully explanatory of the gas kinematic in NGC~5135, but only an approximation useful to understand the overall gas motion. 

\begin{figure*}
\centering
\subfloat[][\emph{\textbf{PV-plot} in \textsf{azim} normalization}]
{\includegraphics[width=1\textwidth]{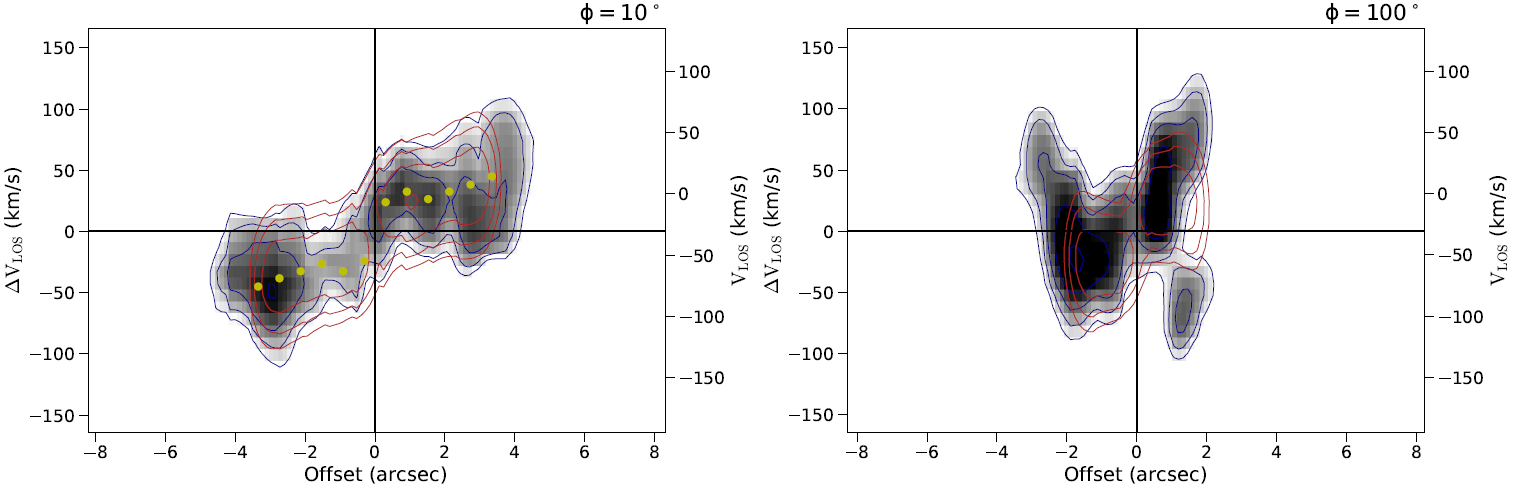}} \quad
\subfloat[][\emph{\textbf{PV-plot} in \textsf{local} normalization}]
{\includegraphics[width=1\textwidth]{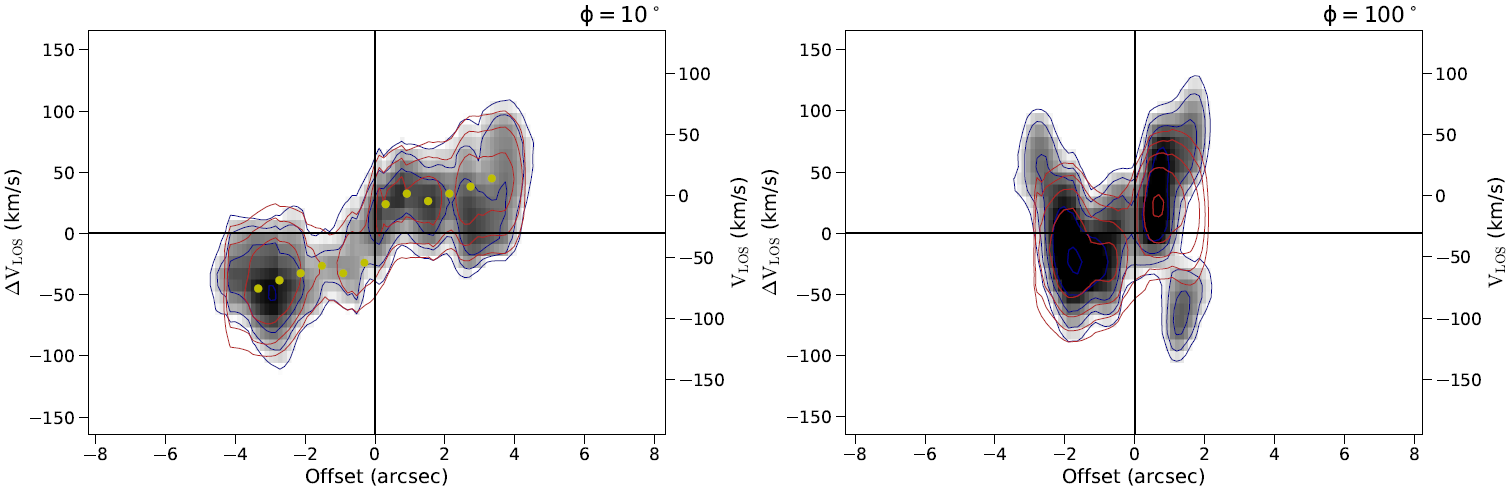}}
\caption{Position-Velocity diagrams along the major-axis ($left$) and the minor-axis ($right$) of the CO(2--1) gas distribution. In the $top$ panels we plot the results of the \textsf{azim} model, 
in the $bottom$ panels those of the \textsf{local} model. Data are shown as grey-scale density levels with overplotted contours (in blue), while the models are
shown as red contours. All contours are shown at $4\sigma$, $16\sigma$ and $32\sigma$. The PA of the major axis is $\phi=10^{\circ}$, 
while that of the minor axis is $\phi=100^{\circ}$. In the P-V diagram along the major axis, the values of the rotational velocity ($V_{rot}$) are shown as yellow filled circles.} 
\label{fig:pvplots}
\end{figure*}
In Figure~\ref{fig:pvplots} we show the position-velocity (P-V) diagrams along the kinematic major ($left$) and minor ($right$) axes of the galaxy disk of NGC~5135 obtained with {\tt$^{\rm 3D}$BAROLO}. 
In the presence of a purely rotating disk, the P-V along the minor axis is expected to be flat, while it assumes an ``S'' shape in presence of radial motions (e.g., \citealt{fraternali2002}). 
Although the P-V diagram along the major axis is similar to what we would expect for a regular spiral galaxy described by a rotating-disk model, the gas distribution along the minor axis 
shows three peculiar regions that cannot be described by the same model, even including a radial velocity component. Note that in NGC~5135 we are not only observing the P-V pattern 
expected for radial motions, but also velocities along the minor axis as large as those along the major axis. This is an unusual behaviour in regular spiral galaxies. 
Moreover, although the overall galaxy structure is well explained by the rotation disk plus radial motions model (see P-V along the major axis), there are clearly three velocity regions along the minor axis 
that either of the two models (i.e., {\tt local} and {\tt azim}) are unable to explain. 
This is visible at an offset of $\sim$$+$1.5 arcsec, as a zone totally empty of data that the model would instead predict to be filled. 
The two ``blobs'' above and below the empty region (at $+$1.5 arcsec offset, $+$70 and $-$70 km s$^{-1}$, corresponding to the 
two peaks in the {\it left} panel of Figure~\ref{fig:pvplots}), are likely related to the presence of an expanding  ``superbubble'' (see next Section).
\begin{figure} 
\includegraphics[width=0.66\columnwidth, angle=270]{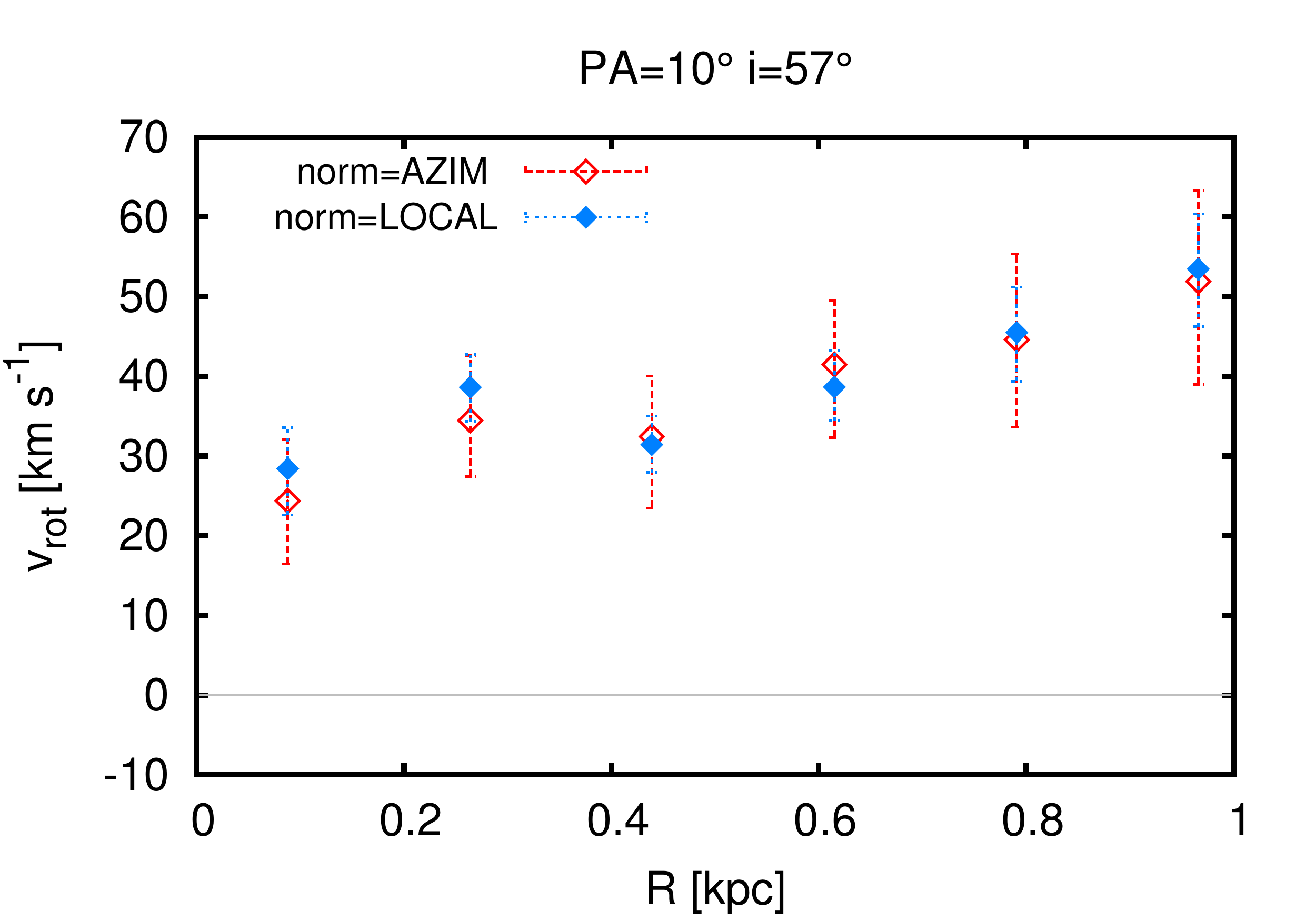}
\includegraphics[width=0.66\columnwidth,angle=270]{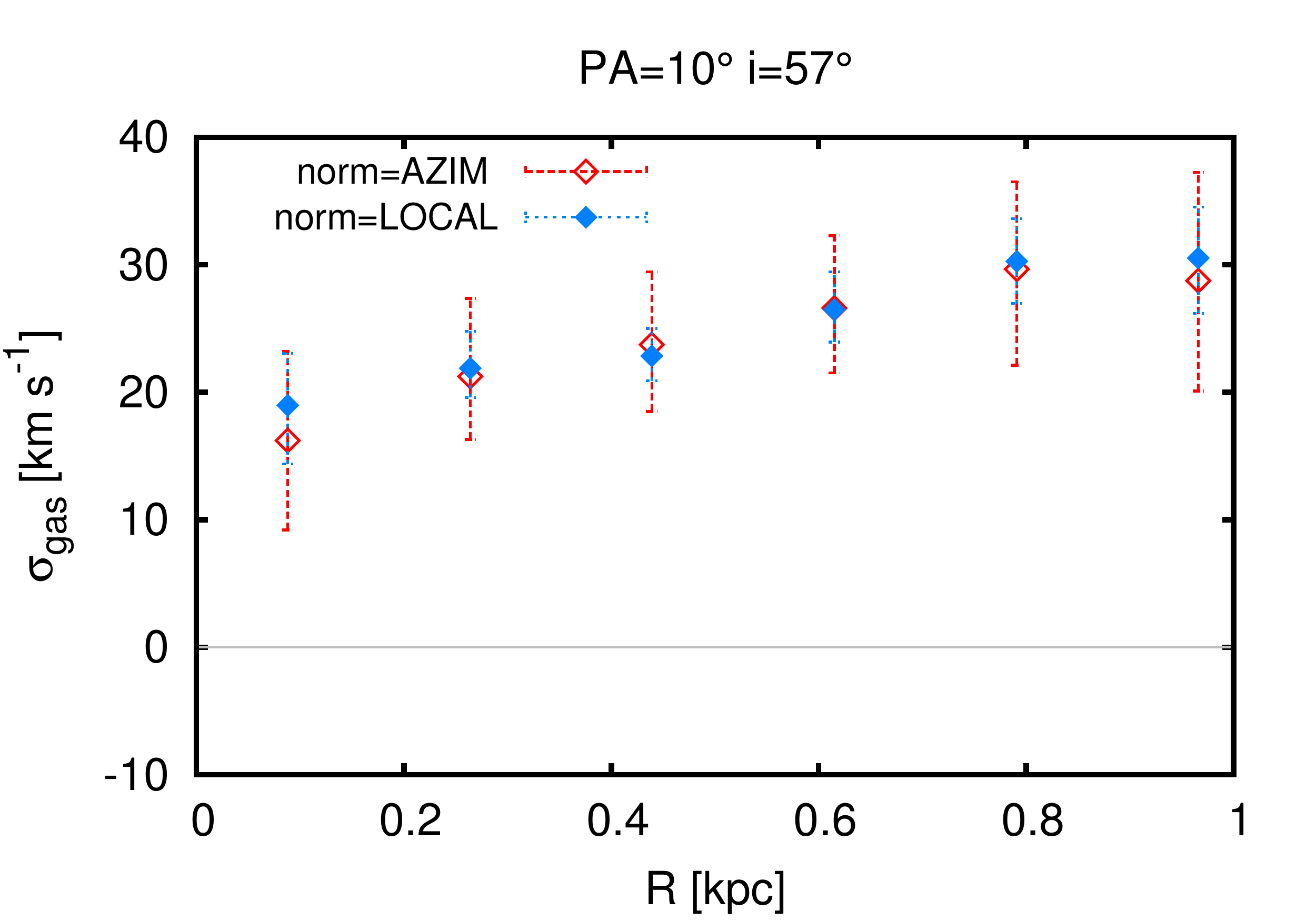}
\includegraphics[width=0.66\columnwidth,angle=270]{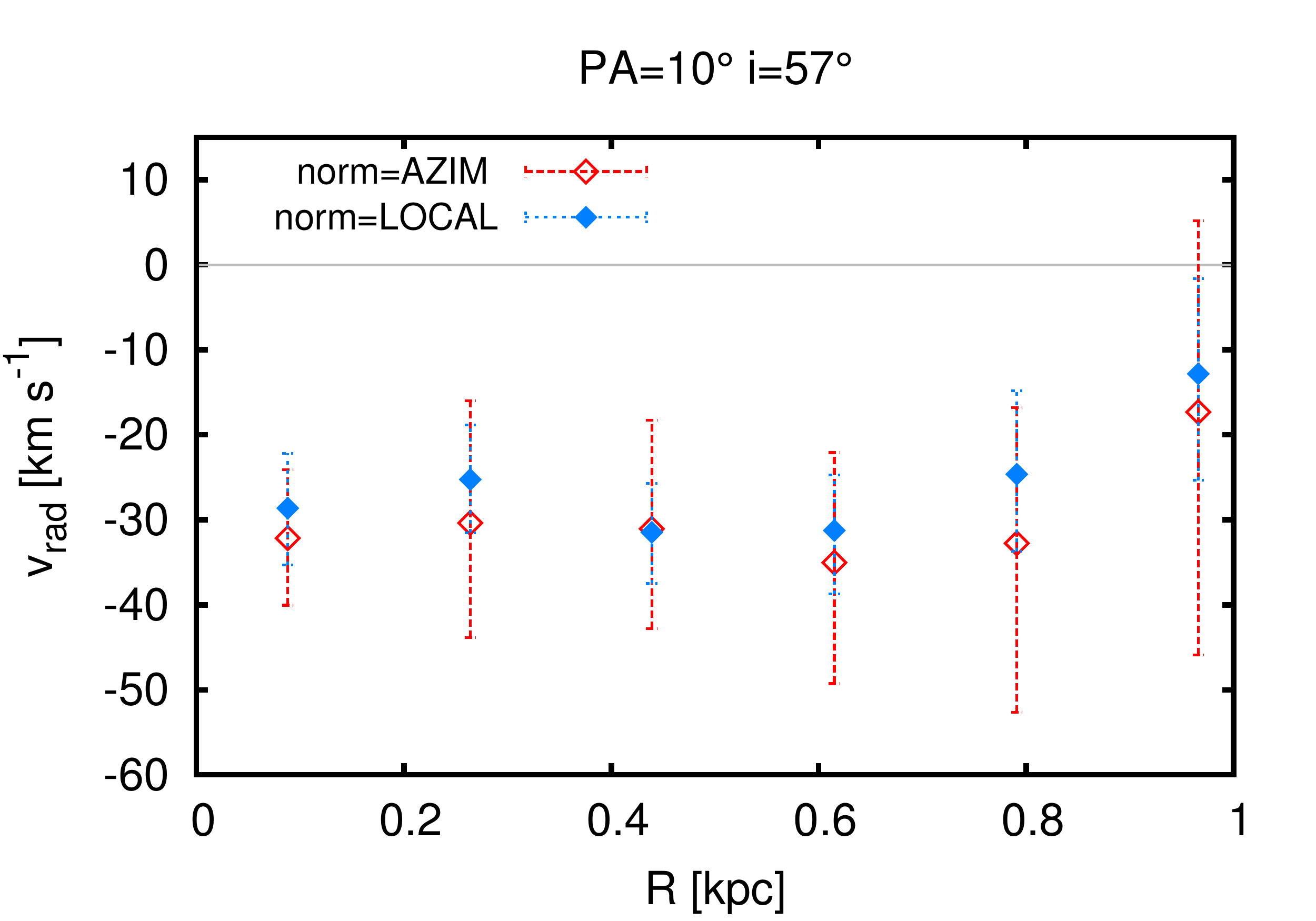}
\caption{CO(2--1) rotation curve of NGC~5135, as derived with {\tt $^{\rm 3D}$BAROLO}: rotation velocity as a function of the distance from the centre ($top$ panel), velocity dispersion (expressed as line width, $central$ panel) and radial velocity ($bottom$ panel). The different colours of the symbols represent the two different models: red open diamonds with error-bars show the results of the {\tt azim} normalisation, while blue filled diamond with error-bars show the results of the {\tt local} model.}
\label{fig:rotcurve1}
\end{figure}
\begin{figure}
\includegraphics[angle=0, width=0.95\columnwidth]{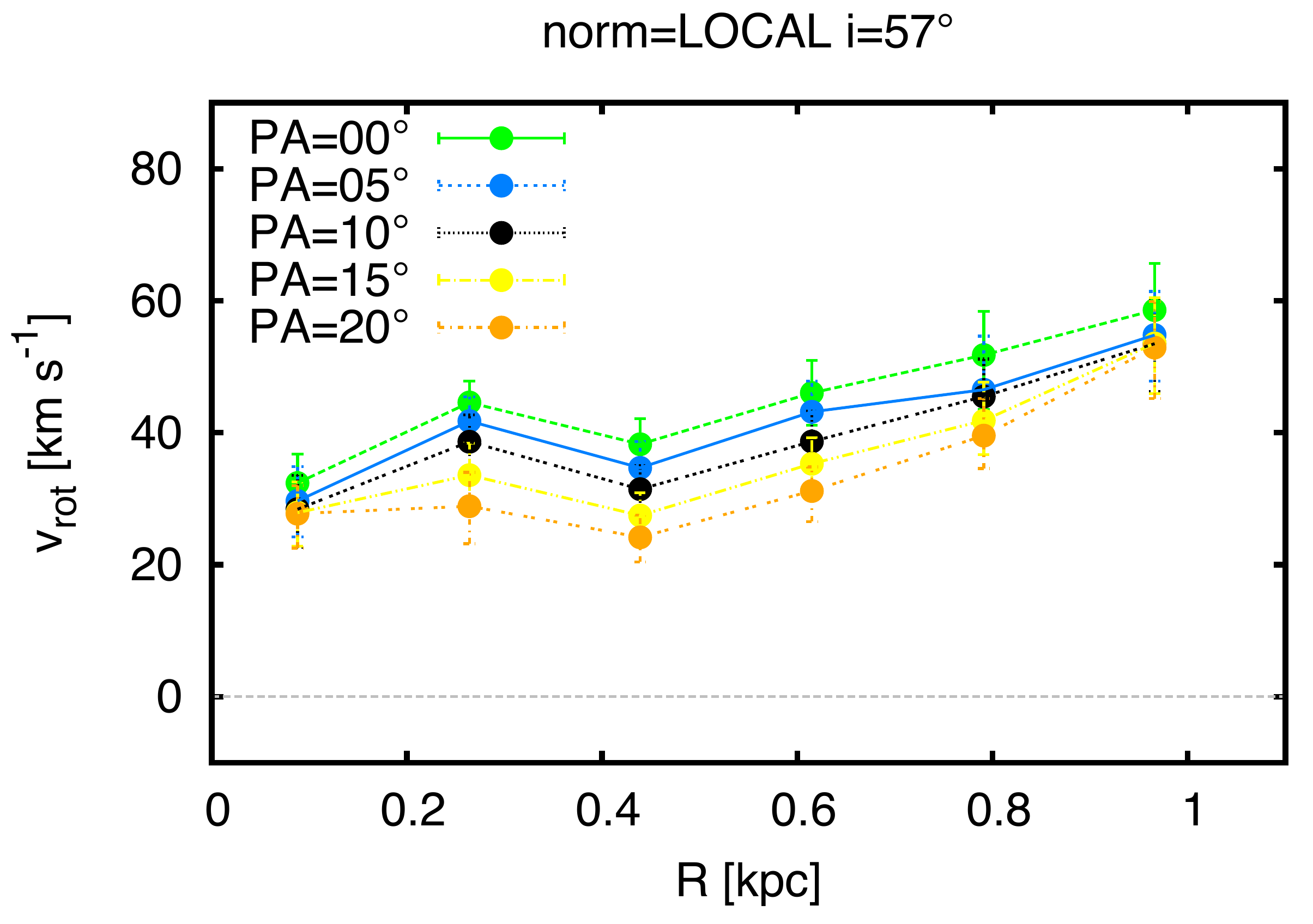}\\
\includegraphics[angle=0, width=0.95\columnwidth]{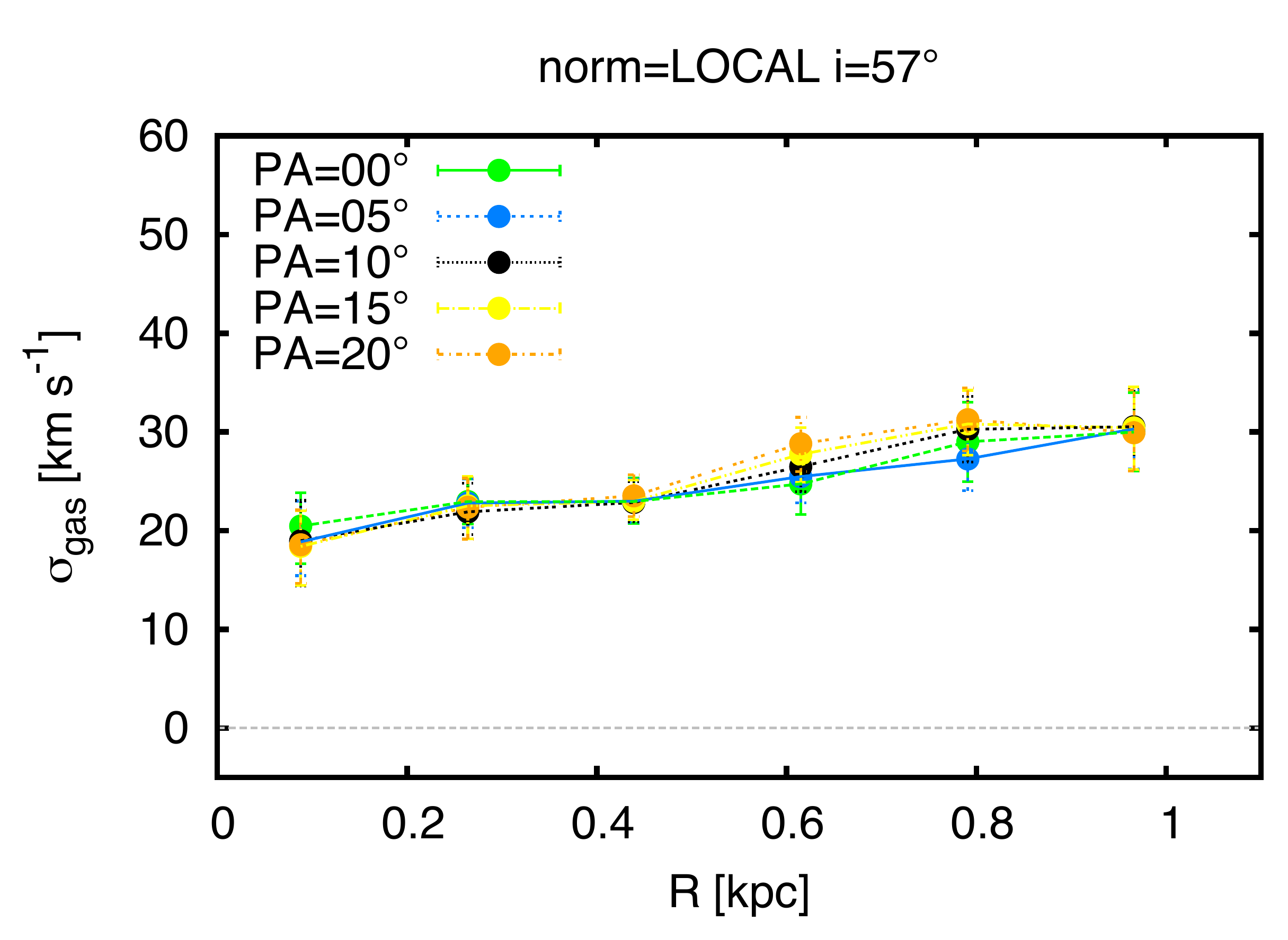}\\
\includegraphics[angle=0, width=0.95\columnwidth]{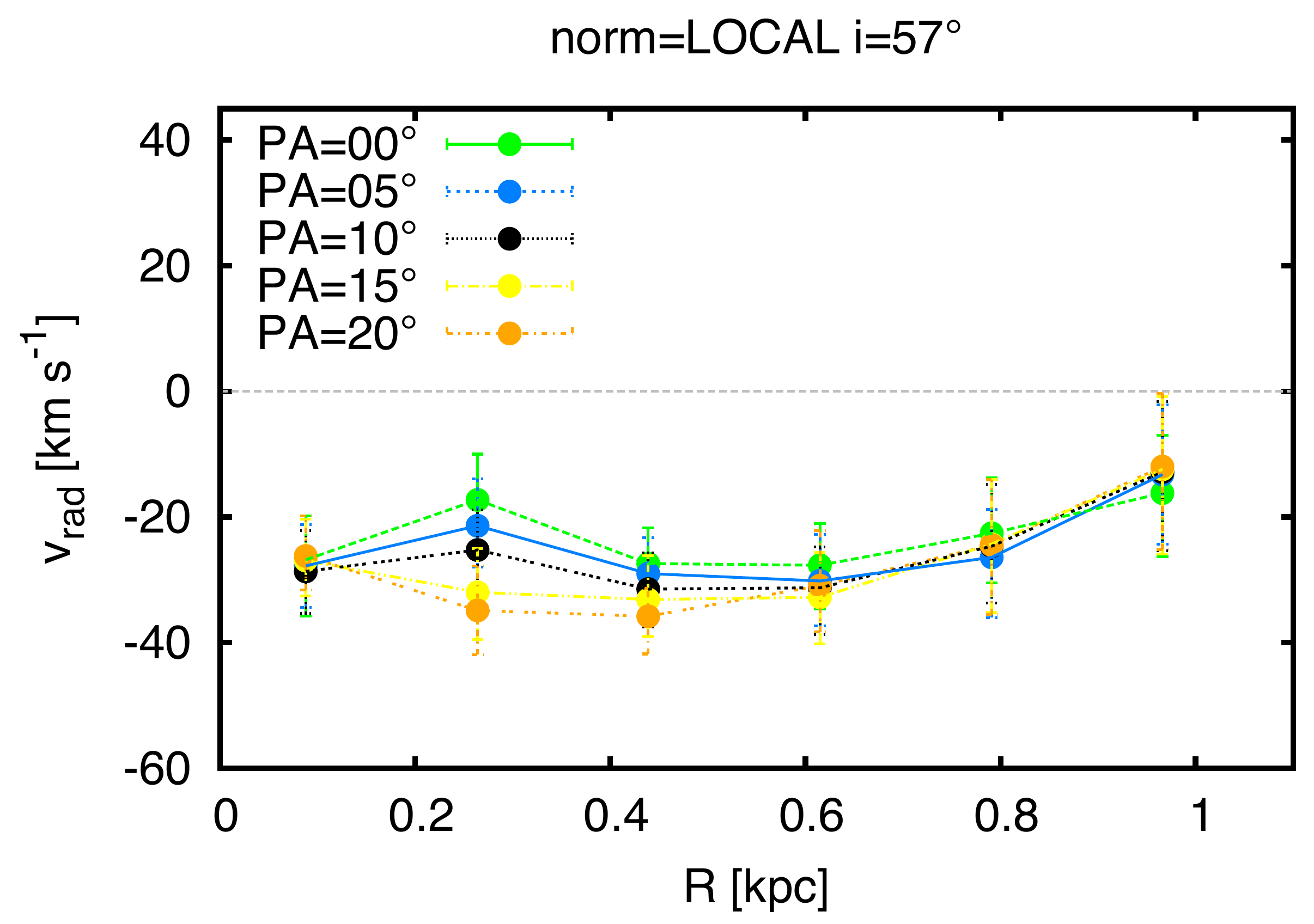}
\caption{Comparison of the CO(2--1) rotation curves, as derived with {\tt $^{\rm 3D}$BAROLO} by considering the {\tt local} normalisation and
different position angles: the different colours of the symbols represent different PA, ranging from $0^{\circ}$ to $20^{\circ}$, in steps of $5^{\circ}$.
In the $top$ panel we show the rotation velocity, in the $central$ panel the velocity dispersion and in the $bottom$ panel the radial velocity, 
as a function of the distance from the centre (in kpc). The grey dotted line identifies the zero-level of $V_{\rm rot}$, $\sigma_{\rm gas}$ and $V_{\rm rad}$.}
\label{fig:rotcurve2}
\end{figure}

In Figure~\ref{fig:rotcurve1}, we show the rotation curve ($V_{\rm rot}$; $top$ panel), the velocity dispersion ($\sigma_{\rm gas}$; $middle$ panel) and the radial velocity ($V_{\rm rad}$; $bottom$ panel) reconstructed by {\tt$^{3D}$BAROLO} as a function of the distance from the galaxy center (up to ~1 kpc). 
The stability of our results is confirmed by the good agreement between the solutions of the two {\tt$^{3D}$BAROLO} models, which are perfectly consistent within the error-bars.
In Figure~\ref{fig:rotcurve2}, the same parameters of Figure~\ref{fig:rotcurve1} are plotted (as derived from the {\tt local} model) 
for different values of the position angle (i.e., varying the PA by steps of 5$^{\circ}$ within the error, around the best-fit value, i.e., $\phi$$=$10$^{\circ}$): 
this shows how robust our results are with respect to changing the PA value. Although the major axis is extracted along the PA, and the minor axis along PA$+$90$^{\circ}$,
the P-V diagrams, as well as the derived velocities, are almost unaffected by these changes. 
From the resulting rotation curve ($top$), we observe an increasing internal rotation of the molecular gas, passing from 20 km s$^{-1}$ to $\sim$50 km s$^{-1}$
at $\sim$1 kpc from the centre, likely to increase further in the outer parts of the galaxy. This is not an obvious results, since it shows a significant (increasing) rotation of the molecular gas 
also in the very inner parts of the galaxy, expected to be ``flat'' and lower in typical nearby spiral galaxies (e.g., \citealt{frank2016}).  
Note that the presence of the bar in NGC~5135 could cause an underestimation of the rotation velocity: since the gas is streaming along the bar, observed velocities are close to rigid-body motion of the bar potential, and hence slower than circular velocity (e.g., \citealt{sofue2016}).\\
The velocity dispersion ($middle$ panel) shows a modest increase with increasing distance, passing from $\simeq$20 to 30 km s$^{-1}$, while the radial velocity ($bottom$) keeps almost constant, 
at around $-$30 km s$^{-1}$, from the galaxy center to $\sim$1 kpc. We stress again that the presence of the bar can influence also the derived $V_{\rm rad}$, whose significantly high values can be due
to large non-circular motions of the gas inside the bar.

\section{The possible origins of the complex gas kinematic in NGC~5135}\label{sec:discussion}
As discussed in the previous section, the overall structure of the molecular gas CO(2--1) in NGC~5135, its density and kinematics can be well explained by a model considering a rotating 
disk perturbed by radial motions. The direction of the radial flow of gas (i.e., in- or out-flow) can be estimated if the closer spiral arm is known, based on the inclination of the disk. 
Assuming that the arms are trailing (i.e., the southern arm is receeding), we find that the molecular gas should flow towards the center of the galaxy: we are facing a gas inflow
towards the BH (for this reason the radial velocity in Figure~\ref{fig:rotcurve1} is negative). 
However, the gas structure is not properly regular, as there are several regions that fall outside the rotating disk $+$ gas inflow model explanation.\\
In Figure~\ref{fig:ALMAimages} we have identified these regions, plus others that would deserve further explanation (marking them with letters and - if needed - encircling with ellipses), for a detailed discussion on the origin and nature of each of them.

Previous observations with different instruments (e.g., HST, VLT-SINFONI; e.g., \citealt{gonzalezdelgado1998, alonsoherrero2006, bedregal2009, colina2012}) have identified the 
central AGN (marked with the ``{\tt A}'' letter), few SF regions at the left edge of the 
inner bar (``{\tt C}'', ``{\tt D}'', ``{\tt E}''), and a SNR in the Southern arm of the inner spiral gas structure (``{\tt G}''). 
The AGN region (``{\tt A}'') is observable in continuum (both B6 and B9), but not in CO line, 
in agreement with the finding that the BH is strongly obscured (i.e., Compton-Thick, CT, from X-ray data; \citealt{levenson2004}): the cold toroidal dust distribution 
has the size of the beam in ALMA B6, thus being unresolved at a linear scale of $\simeq$56 pc. 
The SF regions (``{\tt C}'', ``{\tt D}'', ``{\tt E}'') are clearly identifiable in the data in continuum, CO (both (2--1) and (6--5)) and CS emission. 
Regions ``{\tt B}'' and ``{\tt F}'' were not previously identified in any other bands, and seem to trace a shock front at the right edge of the bar, with an expanding
super-bubble (see below) and gas likely thrown out of the galaxy at high velocity (in two opposite directions, thus showing a very large, double peaked width of the CO line). 
The alignment between zones ``{\tt A}'', ``{\tt B}'' and ``{\tt C}'' seems to indicate the presence of a bar, with enhanced SF at its edges. 
Zone ``{\tt B}'' is likely to trace a shock front between the AGN and zone ``{\tt F}'' (see further explanation), with the latter identifying the two opposite velocities region 
(clearly visible also in Figure~\ref{fig:pvplots} along the minor axis at $\sim$1.5 arcsec and $\pm$70 km s$^{-1}$) at the edge of the bar edge.
region ``{\tt F}'' lies just outside the bar, close to the shock front and is the more puzzling region observed in these data (a tentative explanation is given in the next paragraph).

\begin{figure*}
\includegraphics[width=\textwidth]{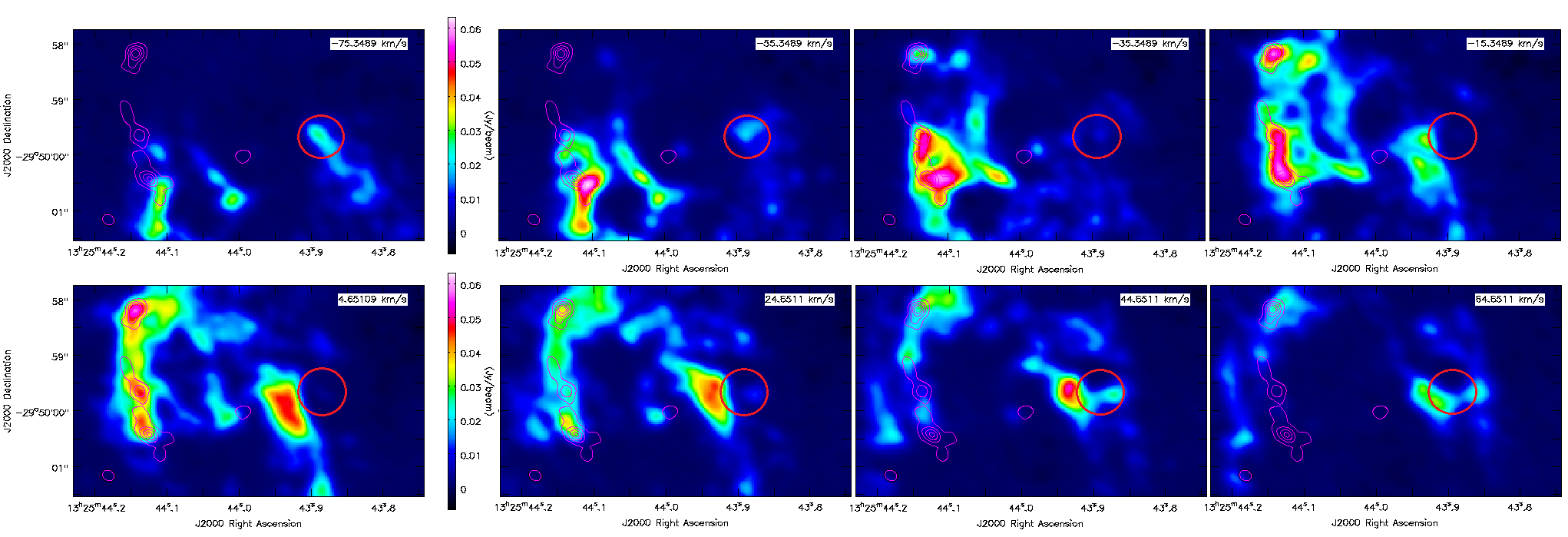}
\caption{Different CO(2--1) velocity channels showing in sequence the appearance -- within the same region -- of: a gas cloud moving at negative velocity, 
a shock front and an empty region (i.e., the ``super-bubble'', encircled in red) peaking at 0 km s$^{-1}$, a gas cloud moving at positive velocity, almost opposite to the first one.}
\label{fig:bubble}
\end{figure*}
Both zones (``{\tt F}'' and ``{\tt H}'') were not identified by any previous observations in other bands. \\
Since we have chosen to be conservative and limit our moment analysis to the inner 4 arcsec ($\sim$1.167 kpc from the center), where the {\tt $^{3D}$BAROLO}
software provides more secure results, region ``{\tt H}'' is out of the area considered for the fit. 
Indeed, the channel-by-channel plot (i.e., Figure~\ref{fig:modelli}, extending to 6 arcsec, $\sim$1.7 kpc) shows that the CO(2--1) distribution in this region can be only partially reproduced, 
with a tail at $\lsimeq$$-$40 km s$^{-1}$ remaining unfitted. 
Our interpretation of region ``{\tt H}'' is that here there could likely be an overlap of two gas components: one following the southern spiral arm motion, the other 
flowing in opposite direction (e.g., the second peak at about $-$80 km s$^{-1}$ in the spectrum shown in Figure~\ref{fig:mom0_spec}). 
The gas component following the arm flow is fitted by the model, the material flowing in the other direction is not.

Regarding region ``{\tt F}'', local emission from a SNR could be a simple explanation, but, differently from zone ``{\tt G}'', in ``{\tt F}'' we do not observe significant emission 
either in continuum nor in spectral lines, but only a very large CO(2--1) velocity dispersion (i.e., produced by the presence of the two opposite velocity peaks, from two gas clouds 
spatially overlapping). Moreover, a SNR in this region cannot be confirmed by data in other bands, as in the case of region ``{\tt G}''.
Hence the SNR origin hypothesis is likely to be discarded for no clear evidence of the presence of the remnant.
The possibility of an outflow from the AGN not lying along the bar (i.e., perpendicular to the galaxy plane and just projected in zone ``{\tt F}'') could be considered, but in this case  
the origin of the shock front in ``{\tt B}'' will be unrelated to the outflow and unclear. Moreover, an outflow of material due to the AGN is also not clearly justified by any 
other evidences, and it is very unlikely that, by chance, the projected position of the out-flowing material along a jet would fall exactly at the edge of the bar. Finally, the outflow 
interpretation cannot explain the material expelled in two opposite velocities in ``{\tt F}''.
Our favourite interpretation is that in ``{\tt F}'' we are facing a flow of material towards the AGN: an in-flow of gas along the bar (connecting the SF region ``{\tt C}'', the AGN ``{\tt A}'' 
and the SF region ``{\tt B}'') could justify the presence of the shock front (``{\tt B}'').
The gas, in-falling through the inner arms by spiralling towards the central BH, could suddenly change its direction as approaching the nucleus, to start following the bar structure: such a flow is 
likely described by the continuum and CO(2--1) velocity pattern observed in the continuum and moment maps (see $left$ panels of  Figures~\ref{fig:ALMAimages} and \ref{fig:ALMAmom}). 

According to our interpretation, the observed high velocity dispersion region in zone ``{\tt F}'' associated to
the zone totally empty of gas between the two opposite velocity clouds (that the {\tt $^{\rm 3D}$BAROLO} model would instead predict to be filled) 
resembles an expanding super-bubble. Super-bubbles are known to be associated to very massive stars (OB-associations): strong stellar winds and subsequent 
SN explosions from those stars inject energy and mass into the ambient ISM, creating shock fronts that sweep-up the ISM (e.g., \citealt{castor1975}). 
As the SN explosions start occurring within the cavity formed by the stellar wind bubbles, super-bubbles are created (e.g., \citealt{mccray1987})
which may eventually reach kiloparsec sizes. These explosions never form a visible SNR, but instead expend their energy in the hot interior as sound waves. 
Both stellar winds and stellar explosions thus power the expansion of the super-bubble in the ISM.
The interstellar gas swept up by super-bubbles generally cools, forming a dense shell around the cavity (observed in HI and H$\alpha$).  
The bubble interior contains hot ($>$10$^6$  K), rarefied material, usually associated with extended diffuse X-ray emission 
(thus appearing as an empty cavity in cold and molecular gas).
In this interpretation the two blobs of gas above and below the observed empty cavity in region ``{\tt F}'' (at $+$1.5 arcsec offset, with velocities of $+$70 and $-$70 km s$^{-1}$, 
corresponding to the two peaks in Figure~\ref{fig:mom0_spec}), 
can be easily explained as gas surrounding the ``super-bubble'' being thrown away in different directions by the expanding spherical shell 
(see \citealt{kamphuis1991,boomsma2008} for detected velocities up to 100 km s$^{-1}$ associated to ``holes'' in HI).
In Figure~\ref{fig:bubble} we show different CO(2--1) velocity channels, where the ``super-bubble'', surrounded by material moving in opposite directions (e.g., peaks at -70 km s$^{-1}$ and  70 km s$^{-1}$, not
present at the systemic velocities), shows up as an empty area (encircled in the plot), with a clear shock front at its left hand side, peaking around the systemic velocities ($\simeq$0 km s$^{-1}$).

In Table~\ref{tab:zones} we list the different zones identified by our data in the inner regions of NGC~5135 and summarise their main characteristics and our interpretation.
\begin{table*}
\caption{Principal zones identified by ALMA in the nuclear region of NGC~5135}
\label{tab:zones}
\begin{tabular}{cll}
\hline
Zone & Characteristics & Possible nature \\ \hline
{\tt A} & central, unresolved at $<$56 pc, continuum only & AGN dusty torus, also studied by \citet{bedregal2009} \\
{\tt B} & high moment 0 CO(2--1), at the edge of the bar & shock front, enhanced SF \\
{\tt C} & high continuum and CO mom 0 (both B6 and B9) & SF region, also studied by \citet{bedregal2009} \\
{\tt D} & high continuum and CO mom 0 (both B6 and B9) & SF region, also studied by \citet{bedregal2009} \\
{\tt E} & high continuum and CO mom 0 (both B6 and B9), but also CS(5--4) & SF region, likely denser than {\tt C, D} \\
{\tt F} & high mom 1 and 2 of CO(2--1), opposite velocity clouds$+$empty zone & expanding ``super-bubble'', not observed before \\
{\tt G} & high continuum and CO mom 0 (both B6 and B9) & SNR also studied by \citet{colina2012}  \\
{\tt H} & high mom 1 and 2 of CO(2--1) &  material flowing opposite to the arm, not observed before\\
\hline
\end{tabular}
\end{table*}

The complex kinematics observed in the central kpc of NGC~5135 is also
intriguing when compared to the most recent models attempting to
describe the dynamics and SF in the central molecular zone
of the Milky Way. In particular, the models described by
\citet{Kruijssen2014,Kruijssen2015,Krumholz2016} predict that the gas is
inflowing within the inner Lindblad resonance, where acoustic
instabilities are stable to gravitational collapse and drive turbulence
in the gas, maintaining large line-widths. As the gas streams in, the
rotation curve passes from flat to solid body. In this region
a substantial quantity of gas is accumulated, the shear drops and the
turbulence is dissipated, making the gas prone to gravitational instability.
In the nuclear regions of NGC~5135 we are likely facing radial motions,
with the gas streaming from kpc scales to $\simeq200\,\mathrm{pc}$.
At scales smaller than 200 pc, the {\tt$^{\rm 3D}$BAROLO} rotation curve
is steep, then it flattens at 200$\lsimeq$R$\lsimeq$400 pc, to then steepen again
towards the outer regions (at least up to $\sim$1 kpc). 
However, within 200 pc, both the {\tt azim} and the
{\tt local} models of {\tt$^{\rm 3D}$BAROLO} find a minimum value
for the CO(2--1) velocity dispersion (see Figure~\ref{fig:rotcurve1}), 
which then increases at larger distances (due to the presence of the SF regions 
the SNR and the super-bubble, observed at the outer edge of this ``ring'').
This, in analogy with what happens in our Galaxy, might suggest that 
in NGC~5135 the gas starts accumulating and becoming 
gravitationally unstable at a scale of $\simeq$200-300 pc from the nucleus.

Although extremely interesting and complex, the picture of the inner regions of NGC 5135 emerging from the current ALMA data is still incomplete: 
the dusty torus is unresolved in continuum and undetected in CO, and the CO velocity maps raise the issue of what generates the two (opposite) high velocity 
dispersion components at the nuclear bar edge and the possible ``super-bubble''. This is a totally new result, needing further investigation.  
To resolve the cold dust (at $<$10--20 pc, possibly at clump level) of the toroidal distribution surrounding and fuelling the AGN at the centre 
of the nearby Seyfert NGC~5135, measuring its Rayleigh-Jeans tail and testing the current models, could be obtained only with higher resolution observations 
(e.g., $\lsimeq$0.03$^{\prime\prime}$--0.05$^{\prime\prime}$ in continuum -- $\sim$5--7 times better than current), 
both in continuum and line, allowing us to spatially resolve the torus and to trace its molecular structure with high density tracers. 
Such high resolution would also allow us to study in detail the complex kinematic of the high velocity dispersion region, in order to spatially resolve the zone with 
opposite velocities, investigating the nature of  the material flowing in or out the nuclear region, and the possible presence of the super-bubble.

\section{Conclusions}
In order to investigate the mutual interplay between AGN and SF activity within galaxies, their relative role in outflow or inflow phenomena, that can be related to 
either negative feedback towards SF or AGN fuelling with enhanced SF, we have analised the ALMA archival data in B6 and B9 for the nearby Seyfert galaxy
NGC~5135. Continuum and line maps have been obtained, showing a complex gas distribution and kinematics. We have modelled the data with a
code fitting the line data cubes of rotating galaxies with a 3D tilted-rings model ({\tt $^{\rm 3D}$BAROLO}, \citealt{diteodoro2015}),
in order to obtain a clearer picture of the phenomena occurring in the inner regions of our target.
Several possible explanations for the complex geometry and gas dynamics in this galaxy have been discussed, with the main result
being the discovery of radial motions overimposed to the galaxy disk rotation, with clear evidence of them consisting of gas inflow towards the nucleus,
occurring through inner spiral arms and a bar. Several regions of enhanced SF due to shocks, outflows and super-bubbles have been identified, clear
signs of a turbulent activity near the galaxy centre and the AGN.

The main results of this work can be here summarised as follows:
\begin{itemize}
\item the ALMA B6 and B9 continuum and CO line data show a complex morphology and gas kinematics in the inner regions of the nearby Seyfert galaxy NGC~5135;
\item at the locus of the central AGN (previously observed with VLT-SINFONI and in X-ray) we detect unresolved flux in both ALMA bands, at a galaxy scale of $<$59 pc in the B6 continuum, 
while do not detected any CO emission, in agreement with the AGN being heavily obscured (as from X-ray). The unresolved flux could contain a possible contribution
from circum-nuclear SF: only higher resolution data will allow us to derive the AGN contribution only;
\item the fluxes measured at $\sim$0.45 and 1.3 mm in the central region, considered an upper limit to the AGN dusty torus emission due to the possible SF contamination,
largely exceed the SED decomposition expectations for torus only (by about two orders of magnitude). If the flux is dominated by the AGN (with negligible contribution from circum-nuclear
SF), this result will imply a significantly higher contribution from cold dust heated by the AGN than predicted by the torus models;
\item the inner regions of the galaxy observed with ALMA show a barred spiral-like structure with two arms and a bar, connecting the central AGN with two enhanced SF regions at the edges
of the bar itself;
\item the analysis of the CO(2-1) moment maps (e.g., gas distribution, velocity field, velocity dispersion) reveals very complex structures and gas kinematics, part of which we tried to explain by rotating disk plus radial motions with turbulence at the edge of the bar;
\item the regions not explainable by the kinematic model identify potential gas outflows/inflows from/towards the AGN, associated to enhanced SF activity, a shock front and 
an expanding super-bubble.     
\end{itemize}

Given its complex structure and the multiple components shown by the ALMA observations of the innermost regions of NGC~5135, this galaxy constitutes a crucial benchmark for 
understanding the feeding and the feedback mechanisms associated to SMBHs. New observations with higher resolution will be crucial to understand the physics and the geometry of the inner AGN region. In a forthcoming paper we will analise in detail the physical properties of the different structures identified in this work, such as the mass flow rate
and the energy associated to the gas inflow and outflow, investigating the role of these regions and of the related processes in the feedback or the enhancing of SF, 
and in feeding the BH (e.g., the rate of gas flow towards the AGN).

\section*{Acknowledgments}
The authors wish to thank F. Fraternali and E. Di Teodoro for their great help in the use of the {\tt $^{\rm 3D}$BAROLO} software and for very fruitful discussions about the interpretation of 
the results of this work.
This paper makes use of the following ALMA data: 2013.1.00243.S  and 2013.1.00581. ALMA is a partnership of ESO (representing its member states), NSF (USA) and NINS (Japan), together with NRC (Canada), NSC and ASIAA (Taiwan), and KASI (Republic of Korea), in cooperation with the Republic of Chile. The Joint ALMA Observatory is operated by ESO, AUI/NRAO and NAOJ.
CG acknowledges funding from the INAF PRIN-SKA 2017 program 1.05.01.88.04.
MM acknowledges partial financial support by the Italian {\it Ministero dell'Istruzione, Universit\`a e Ricerca} through the grant {\it Progetti Premiali 2012--iALMA} (CUP C52I13000140001). 

\bibliographystyle{mnras}
\bibliography{mybib}

\begin{thebibliography}{}
\makeatletter
\relax
\def\mn@urlcharsother{\let\do\@makeother \do\$\do\&\do\#\do\^\do\_\do\%\do\~}
\def\mn@doi{\begingroup\mn@urlcharsother \@ifnextchar [ {\mn@doi@}
  {\mn@doi@[]}}
\def\mn@doi@[#1]#2{\def\@tempa{#1}\ifx\@tempa\@empty \href
  {http://dx.doi.org/#2} {doi:#2}\else \href {http://dx.doi.org/#2} {#1}\fi
  \endgroup}
\def\mn@eprint#1#2{\mn@eprint@#1:#2::\@nil}
\def\mn@eprint@arXiv#1{\href {http://arxiv.org/abs/#1} {{\tt arXiv:#1}}}
\def\mn@eprint@dblp#1{\href {http://dblp.uni-trier.de/rec/bibtex/#1.xml}
  {dblp:#1}}
\def\mn@eprint@#1:#2:#3:#4\@nil{\def\@tempa {#1}\def\@tempb {#2}\def\@tempc
  {#3}\ifx \@tempc \@empty \let \@tempc \@tempb \let \@tempb \@tempa \fi \ifx
  \@tempb \@empty \def\@tempb {arXiv}\fi \@ifundefined
  {mn@eprint@\@tempb}{\@tempb:\@tempc}{\expandafter \expandafter \csname
  mn@eprint@\@tempb\endcsname \expandafter{\@tempc}}}

\bibitem[\protect\citeauthoryear{{Alonso-Herrero}, {Rieke}, {Rieke}, {Colina},
  {P{\'e}rez-Gonz{\'a}lez}  \& {Ryder}}{{Alonso-Herrero}
  et~al.}{2006}]{alonsoherrero2006}
{Alonso-Herrero} A.,  {Rieke} G.~H.,  {Rieke} M.~J.,  {Colina} L.,
  {P{\'e}rez-Gonz{\'a}lez} P.~G.,   {Ryder} S.~D.,  2006, \mn@doi [\apj]
  {10.1086/506958}, \href {http://adsabs.harvard.edu/abs/2006ApJ...650..835A}
  {650, 835}

\bibitem[\protect\citeauthoryear{{Baldwin}, {Phillips}  \&
  {Terlevich}}{{Baldwin} et~al.}{1981}]{baldwin1981}
{Baldwin} J.~A.,  {Phillips} M.~M.,   {Terlevich} R.,  1981, \mn@doi [\pasp]
  {10.1086/130766}, \href {http://adsabs.harvard.edu/abs/1981PASP...93....5B}
  {93, 5}

\bibitem[\protect\citeauthoryear{{Bedregal}, {Colina}, {Alonso-Herrero}  \&
  {Arribas}}{{Bedregal} et~al.}{2009}]{bedregal2009}
{Bedregal} A.~G.,  {Colina} L.,  {Alonso-Herrero} A.,   {Arribas} S.,  2009,
  \mn@doi [\apj] {10.1088/0004-637X/698/2/1852}, \href
  {http://adsabs.harvard.edu/abs/2009ApJ...698.1852B} {698, 1852}

\bibitem[\protect\citeauthoryear{{Bedregal}, {Colina}, {Azzollini}, {Arribas}
  \& {Alonso-Herrero}}{{Bedregal} et~al.}{2011}]{bedregal2011}
{Bedregal} A.~G.,  {Colina} L.,  {Azzollini} R.,  {Arribas} S.,
  {Alonso-Herrero} A.,  2011, in {Wang} W.,  {Lu} J.,  {Luo} Z.,  {Yang} Z.,
  {Hua} H.,   {Chen} Z.,  eds,  Astronomical Society of the Pacific Conference
  Series Vol. 446, Galaxy Evolution: Infrared to Millimeter Wavelength
  Perspective. p.~83 (\mn@eprint {arXiv} {1102.0012})

\bibitem[\protect\citeauthoryear{{Berta} et~al.,}{{Berta}
  et~al.}{2013}]{berta2013}
{Berta} S.,  et~al., 2013, \mn@doi [\aap] {10.1051/0004-6361/201220859}, \href
  {http://adsabs.harvard.edu/abs/2013A%26A...551A.100B} {551, A100}

\bibitem[\protect\citeauthoryear{{Boomsma}, {Oosterloo}, {Fraternali}, {van der
  Hulst}  \& {Sancisi}}{{Boomsma} et~al.}{2008}]{boomsma2008}
{Boomsma} R.,  {Oosterloo} T.~A.,  {Fraternali} F.,  {van der Hulst} J.~M.,
  {Sancisi} R.,  2008, \mn@doi [\aap] {10.1051/0004-6361:200810120}, \href
  {http://adsabs.harvard.edu/abs/2008A%26A...490..555B} {490, 555}

\bibitem[\protect\citeauthoryear{{Bryant} \& {Scoville}}{{Bryant} \&
  {Scoville}}{1999}]{bryant1999}
{Bryant} P.~M.,  {Scoville} N.~Z.,  1999, \mn@doi [\aj] {10.1086/300879}, \href
  {http://adsabs.harvard.edu/abs/1999AJ....117.2632B} {117, 2632}

\bibitem[\protect\citeauthoryear{{Castor}, {McCray}  \& {Weaver}}{{Castor}
  et~al.}{1975}]{castor1975}
{Castor} J.,  {McCray} R.,   {Weaver} R.,  1975, \mn@doi [\apjl]
  {10.1086/181908}, \href {http://adsabs.harvard.edu/abs/1975ApJ...200L.107C}
  {200, L107}

\bibitem[\protect\citeauthoryear{{Cicone} et~al.,}{{Cicone}
  et~al.}{2014}]{cicone2014}
{Cicone} C.,  et~al., 2014, \mn@doi [\aap] {10.1051/0004-6361/201322464}, \href
  {http://adsabs.harvard.edu/abs/2014A%26A...562A..21C} {562, A21}

\bibitem[\protect\citeauthoryear{{Colina}, {Pereira-Santaella},
  {Alonso-Herrero}, {Bedregal}  \& {Arribas}}{{Colina}
  et~al.}{2012}]{colina2012}
{Colina} L.,  {Pereira-Santaella} M.,  {Alonso-Herrero} A.,  {Bedregal} A.~G.,
   {Arribas} S.,  2012, \mn@doi [\apj] {10.1088/0004-637X/749/2/116}, \href
  {http://adsabs.harvard.edu/abs/2012ApJ...749..116C} {749, 116}

\bibitem[\protect\citeauthoryear{{Combes}}{{Combes}}{2012}]{combes2012}
{Combes} F.,  2012, in Journal of Physics Conference Series. p. 012041
  (\mn@eprint {arXiv} {1111.4770}), \mn@doi{10.1088/1742-6596/372/1/012041}

\bibitem[\protect\citeauthoryear{{Combes} et~al.,}{{Combes}
  et~al.}{2013}]{combes2013}
{Combes} F.,  et~al., 2013, \mn@doi [\aap] {10.1051/0004-6361/201322288}, \href
  {http://adsabs.harvard.edu/abs/2013%A26A...558A.124C} {558, A124}

\bibitem[\protect\citeauthoryear{{Di Teodoro} \& {Fraternali}}{{Di Teodoro} \&
  {Fraternali}}{2015}]{diteodoro2015}
{Di Teodoro} E.~M.,  {Fraternali} F.,  2015, \mn@doi [\mnras]
  {10.1093/mnras/stv1213}, \href
  {http://adsabs.harvard.edu/abs/2015MNRAS.451.3021D} {451, 3021}

\bibitem[\protect\citeauthoryear{{Downes} \& {Solomon}}{{Downes} \&
  {Solomon}}{1998}]{downes1998}
{Downes} D.,  {Solomon} P.~M.,  1998, \mn@doi [\apj] {10.1086/306339}, \href
  {http://adsabs.harvard.edu/abs/1998ApJ...507..615D} {507, 615}

\bibitem[\protect\citeauthoryear{{Fabian}}{{Fabian}}{2012}]{fabian2012}
{Fabian} A.~C.,  2012, \mn@doi [\araa] {10.1146/annurev-astro-081811-125521},
  \href {http://adsabs.harvard.edu/abs/2012ARA%26A..50..455F} {50, 455}

\bibitem[\protect\citeauthoryear{{Feruglio}, {Maiolino}, {Piconcelli}, {Menci},
  {Aussel}, {Lamastra}  \& {Fiore}}{{Feruglio} et~al.}{2010}]{feruglio2010}
{Feruglio} C.,  {Maiolino} R.,  {Piconcelli} E.,  {Menci} N.,  {Aussel} H.,
  {Lamastra} A.,   {Fiore} F.,  2010, \mn@doi [\aap]
  {10.1051/0004-6361/201015164}, \href
  {http://adsabs.harvard.edu/abs/2010A%26A...518L.155F} {518, L155}

\bibitem[\protect\citeauthoryear{{Frank}, {de Blok}, {Walter}, {Leroy}  \&
  {Carignan}}{{Frank} et~al.}{2016}]{frank2016}
{Frank} B.~S.,  {de Blok} W.~J.~G.,  {Walter} F.,  {Leroy} A.,   {Carignan} C.,
   2016, \mn@doi [\aj] {10.3847/0004-6256/151/4/94}, \href
  {http://adsabs.harvard.edu/abs/2016AJ....151...94F} {151, 94}

\bibitem[\protect\citeauthoryear{{Fraternali}, {van Moorsel}, {Sancisi}  \&
  {Oosterloo}}{{Fraternali} et~al.}{2002}]{fraternali2002}
{Fraternali} F.,  {van Moorsel} G.,  {Sancisi} R.,   {Oosterloo} T.,  2002,
  \mn@doi [\aj] {10.1086/340358}, \href
  {http://adsabs.harvard.edu/abs/2002AJ....123.3124F} {123, 3124}

\bibitem[\protect\citeauthoryear{{Garc{\'{\i}}a-Burillo}
  et~al.,}{{Garc{\'{\i}}a-Burillo} et~al.}{2014}]{garciaburillo2014}
{Garc{\'{\i}}a-Burillo} S.,  et~al., 2014, \mn@doi [\aap]
  {10.1051/0004-6361/201423843}, \href
  {http://adsabs.harvard.edu/abs/2014A%26A...567A.125G} {567, A125}

\bibitem[\protect\citeauthoryear{{Garc{\'{\i}}a-Burillo}
  et~al.,}{{Garc{\'{\i}}a-Burillo} et~al.}{2016}]{garciaburillo2016}
{Garc{\'{\i}}a-Burillo} S.,  et~al., 2016, \mn@doi [\apjl]
  {10.3847/2041-8205/823/1/L12}, \href
  {http://adsabs.harvard.edu/abs/2016ApJ...823L..12G} {823, L12}

\bibitem[\protect\citeauthoryear{{Gonz{\'a}lez Delgado}, {Heckman},
  {Leitherer}, {Meurer}, {Krolik}, {Wilson}, {Kinney}  \&
  {Koratkar}}{{Gonz{\'a}lez Delgado} et~al.}{1998}]{gonzalezdelgado1998}
{Gonz{\'a}lez Delgado} R.~M.,  {Heckman} T.,  {Leitherer} C.,  {Meurer} G.,
  {Krolik} J.,  {Wilson} A.~S.,  {Kinney} A.,   {Koratkar} A.,  1998, \mn@doi
  [\apj] {10.1086/306154}, \href
  {http://adsabs.harvard.edu/abs/1998ApJ...505..174G} {505, 174}

\bibitem[\protect\citeauthoryear{{Gooch}}{{Gooch}}{1996}]{gooch1996}
{Gooch} R.,  1996, in {Jacoby} G.~H.,  {Barnes} J.,  eds,  Astronomical Society
  of the Pacific Conference Series Vol. 101, Astronomical Data Analysis
  Software and Systems V. p.~80

\bibitem[\protect\citeauthoryear{{Gruppioni} et~al.,}{{Gruppioni}
  et~al.}{2016}]{gruppioni2016}
{Gruppioni} C.,  et~al., 2016, \mn@doi [\mnras] {10.1093/mnras/stw577}, \href
  {http://adsabs.harvard.edu/abs/2016MNRAS.458.4297G} {458, 4297}

\bibitem[\protect\citeauthoryear{{Harrison} et~al.,}{{Harrison}
  et~al.}{2016}]{harrison2016}
{Harrison} C.~M.,  et~al., 2016, \mn@doi [\mnras] {10.1093/mnras/stv2727},
  \href {http://adsabs.harvard.edu/abs/2016MNRAS.456.1195H} {456, 1195}

\bibitem[\protect\citeauthoryear{{Hopkins}, {Richards}  \&
  {Hernquist}}{{Hopkins} et~al.}{2007}]{hopkins2007}
{Hopkins} P.~F.,  {Richards} G.~T.,   {Hernquist} L.,  2007, \mn@doi [\apj]
  {10.1086/509629}, \href {http://adsabs.harvard.edu/abs/2007ApJ...654..731H}
  {654, 731}

\bibitem[\protect\citeauthoryear{{Kamphuis}, {Sancisi}  \& {van der
  Hulst}}{{Kamphuis} et~al.}{1991}]{kamphuis1991}
{Kamphuis} J.,  {Sancisi} R.,   {van der Hulst} T.,  1991, \aap, \href
  {http://adsabs.harvard.edu/abs/1991A%26A...244L..29K} {244, L29}

\bibitem[\protect\citeauthoryear{{Kruijssen}, {Longmore}, {Elmegreen},
  {Murray}, {Bally}, {Testi}  \& {Kennicutt}}{{Kruijssen}
  et~al.}{2014}]{Kruijssen2014}
{Kruijssen} J.~M.~D.,  {Longmore} S.~N.,  {Elmegreen} B.~G.,  {Murray} N.,
  {Bally} J.,  {Testi} L.,   {Kennicutt} R.~C.,  2014, \mn@doi [\mnras]
  {10.1093/mnras/stu494}, \href
  {http://adsabs.harvard.edu/abs/2014MNRAS.440.3370K} {440, 3370}

\bibitem[\protect\citeauthoryear{{Kruijssen}, {Dale}  \&
  {Longmore}}{{Kruijssen} et~al.}{2015}]{Kruijssen2015}
{Kruijssen} J.~M.~D.,  {Dale} J.~E.,   {Longmore} S.~N.,  2015, \mn@doi
  [\mnras] {10.1093/mnras/stu2526}, \href
  {http://adsabs.harvard.edu/abs/2015MNRAS.447.1059K} {447, 1059}

\bibitem[\protect\citeauthoryear{{Krumholz} \& {Burkhart}}{{Krumholz} \&
  {Burkhart}}{2016}]{Krumholz2016}
{Krumholz} M.~R.,  {Burkhart} B.,  2016, \mn@doi [\mnras]
  {10.1093/mnras/stw434}, \href
  {http://adsabs.harvard.edu/abs/2016MNRAS.458.1671K} {458, 1671}

\bibitem[\protect\citeauthoryear{{Lapi}, {Raimundo}, {Aversa}, {Cai},
  {Negrello}, {Celotti}, {De Zotti}  \& {Danese}}{{Lapi}
  et~al.}{2014}]{lapi2014}
{Lapi} A.,  {Raimundo} S.,  {Aversa} R.,  {Cai} Z.-Y.,  {Negrello} M.,
  {Celotti} A.,  {De Zotti} G.,   {Danese} L.,  2014, \mn@doi [\apj]
  {10.1088/0004-637X/782/2/69}, \href
  {http://adsabs.harvard.edu/abs/2014ApJ...782...69L} {782, 69}

\bibitem[\protect\citeauthoryear{{Levenson}, {Weaver}, {Heckman}, {Awaki}  \&
  {Terashima}}{{Levenson} et~al.}{2004}]{levenson2004}
{Levenson} N.~A.,  {Weaver} K.~A.,  {Heckman} T.~M.,  {Awaki} H.,   {Terashima}
  Y.,  2004, \mn@doi [\apj] {10.1086/380836}, \href
  {http://adsabs.harvard.edu/abs/2004ApJ...602..135L} {602, 135}

\bibitem[\protect\citeauthoryear{{McCray} \& {Kafatos}}{{McCray} \&
  {Kafatos}}{1987}]{mccray1987}
{McCray} R.,  {Kafatos} M.,  1987, \mn@doi [\apj] {10.1086/165267}, \href
  {http://adsabs.harvard.edu/abs/1987ApJ...317..190M} {317, 190}

\bibitem[\protect\citeauthoryear{{Mulchaey} \& {Regan}}{{Mulchaey} \&
  {Regan}}{1997}]{mulchaey1997}
{Mulchaey} J.~S.,  {Regan} M.~W.,  1997, \mn@doi [\apjl] {10.1086/310710},
  \href {http://adsabs.harvard.edu/abs/1997ApJ...482L.135M} {482, L135}

\bibitem[\protect\citeauthoryear{{Phillips}, {Charles}  \&
  {Baldwin}}{{Phillips} et~al.}{1983}]{phillips1983}
{Phillips} M.~M.,  {Charles} P.~A.,   {Baldwin} J.~A.,  1983, \mn@doi [\apj]
  {10.1086/160797}, \href {http://adsabs.harvard.edu/abs/1983ApJ...266..485P}
  {266, 485}

\bibitem[\protect\citeauthoryear{{Roelofs}}{{Roelofs}}{1995}]{roelofs1995}
{Roelofs} G.,  1995, PhD thesis, THE UNIVERSITY OF CHICAGO.

\bibitem[\protect\citeauthoryear{{Rush}, {Malkan}  \& {Spinoglio}}{{Rush}
  et~al.}{1993}]{rush1993}
{Rush} B.,  {Malkan} M.~A.,   {Spinoglio} L.,  1993, \mn@doi [\apjs]
  {10.1086/191837}, \href {http://adsabs.harvard.edu/abs/1993ApJS...89....1R}
  {89, 1}

\bibitem[\protect\citeauthoryear{{Saito} et~al.,}{{Saito}
  et~al.}{2017}]{saito2017}
{Saito} T.,  et~al., 2017, \mn@doi [\apj] {10.3847/1538-4357/835/2/174}, \href
  {http://adsabs.harvard.edu/abs/2017ApJ...835..174S} {835, 174}

\bibitem[\protect\citeauthoryear{{Sandage}}{{Sandage}}{1978}]{sandage1978}
{Sandage} A.,  1978, \mn@doi [\aj] {10.1086/112271}, \href
  {http://adsabs.harvard.edu/abs/1978AJ.....83..904S} {83, 904}

\bibitem[\protect\citeauthoryear{{Sanders}, {Mazzarella}, {Kim}, {Surace}  \&
  {Soifer}}{{Sanders} et~al.}{2003}]{sanders2003}
{Sanders} D.~B.,  {Mazzarella} J.~M.,  {Kim} D.-C.,  {Surace} J.~A.,   {Soifer}
  B.~T.,  2003, \mn@doi [\aj] {10.1086/376841}, \href
  {http://adsabs.harvard.edu/abs/2003AJ....126.1607S} {126, 1607}

\bibitem[\protect\citeauthoryear{{Schinnerer}, {Eckart}, {Tacconi}, {Genzel}
  \& {Downes}}{{Schinnerer} et~al.}{2000}]{Schinnerer2000}
{Schinnerer} E.,  {Eckart} A.,  {Tacconi} L.~J.,  {Genzel} R.,   {Downes} D.,
  2000, \mn@doi [\apj] {10.1086/308702}, \href
  {http://adsabs.harvard.edu/abs/2000ApJ...533..850S} {533, 850}

\bibitem[\protect\citeauthoryear{{Schoenmakers}, {Franx}  \& {de
  Zeeuw}}{{Schoenmakers} et~al.}{1997}]{schoenmakers1997}
{Schoenmakers} R.~H.~M.,  {Franx} M.,   {de Zeeuw} P.~T.,  1997, \mn@doi
  [\mnras] {10.1093/mnras/292.2.349}, \href
  {http://adsabs.harvard.edu/abs/1997MNRAS.292..349S} {292, 349}

\bibitem[\protect\citeauthoryear{{Shlosman} \& {Heller}}{{Shlosman} \&
  {Heller}}{2002}]{shlosman2002}
{Shlosman} I.,  {Heller} C.~H.,  2002, \mn@doi [\apj] {10.1086/324403}, \href
  {http://adsabs.harvard.edu/abs/2002ApJ...565..921S} {565, 921}

\bibitem[\protect\citeauthoryear{{Sofue}}{{Sofue}}{2016}]{sofue2016}
{Sofue} Y.,  2016, \mn@doi [\pasj] {10.1093/pasj/psv103}, \href
  {http://adsabs.harvard.edu/abs/2016PASJ...68....2S} {68, 2}

\bibitem[\protect\citeauthoryear{{Somerville} \& {Dav{\'e}}}{{Somerville} \&
  {Dav{\'e}}}{2015}]{somerville2015}
{Somerville} R.~S.,  {Dav{\'e}} R.,  2015, \mn@doi [\araa]
  {10.1146/annurev-astro-082812-140951}, \href
  {http://adsabs.harvard.edu/abs/2015ARA%26A..53...51S} {53, 51}

\bibitem[\protect\citeauthoryear{{Swaters}, {Schoenmakers}, {Sancisi}  \& {van
  Albada}}{{Swaters} et~al.}{1999}]{swaters1999}
{Swaters} R.~A.,  {Schoenmakers} R.~H.~M.,  {Sancisi} R.,   {van Albada} T.~S.,
   1999, \mn@doi [\mnras] {10.1046/j.1365-8711.1999.02332.x}, \href
  {http://adsabs.harvard.edu/abs/1999MNRAS.304..330S} {304, 330}

\bibitem[\protect\citeauthoryear{{Tommasin}, {Spinoglio}, {Malkan}, {Smith},
  {Gonz{\'a}lez-Alfonso}  \& {Charmandaris}}{{Tommasin}
  et~al.}{2008}]{tommasin2008}
{Tommasin} S.,  {Spinoglio} L.,  {Malkan} M.~A.,  {Smith} H.,
  {Gonz{\'a}lez-Alfonso} E.,   {Charmandaris} V.,  2008, \mn@doi [\apj]
  {10.1086/527290}, \href {http://adsabs.harvard.edu/abs/2008ApJ...676..836T}
  {676, 836}

\bibitem[\protect\citeauthoryear{{Tommasin}, {Spinoglio}, {Malkan}  \&
  {Fazio}}{{Tommasin} et~al.}{2010}]{tommasin2010}
{Tommasin} S.,  {Spinoglio} L.,  {Malkan} M.~A.,   {Fazio} G.,  2010, \mn@doi
  [\apj] {10.1088/0004-637X/709/2/1257}, \href
  {http://adsabs.harvard.edu/abs/2010ApJ...709.1257T} {709, 1257}

\bibitem[\protect\citeauthoryear{{Ulvestad} \& {Wilson}}{{Ulvestad} \&
  {Wilson}}{1989}]{ulvestad1989}
{Ulvestad} J.~S.,  {Wilson} A.~S.,  1989, \mn@doi [\apj] {10.1086/167737},
  \href {http://adsabs.harvard.edu/abs/1989ApJ...343..659U} {343, 659}

\bibitem[\protect\citeauthoryear{{V{\'e}ron-Cetty} \&
  {V{\'e}ron}}{{V{\'e}ron-Cetty} \& {V{\'e}ron}}{2006}]{veroncetty2006}
{V{\'e}ron-Cetty} M.-P.,  {V{\'e}ron} P.,  2006, \mn@doi [\aap]
  {10.1051/0004-6361:20065177}, \href
  {http://adsabs.harvard.edu/abs/2006A%26A...455..773V} {455, 773}

\makeatother
\end{thebibliography}

\label{lastpage}

\end{document}